# SPARC: Evidence-guided adaptive model composition for materials discovery

Sissi Feng[1,†], Yuchen Han[1,2,†], Runze Zhang[1,2,†], Jiaru Bai[3,4], Yixi Li[1], Brandon Sutherland[1], Ali Shayesteh Zeraati[1], Jason Hattrick-Simpers[1,2], Varinia Bernales[1,3,5], Alán Aspuru-Guzik[1,2,3,4,5,6,7,8,9], Yang Bai[1,*]

[1]*Acceleration Consortium, University of Toronto, Toronto, Ontario, Canada*

[2]*Department of Materials Science and Engineering, University of Toronto, Toronto, Ontario, Canada*

[3]*Department of Chemistry, University of Toronto, Toronto, ON M5S 3H6, Canada*

[4]*Vector Institute for Artificial Intelligence, W1140-108 College St., Schwartz Reisman Innovation Campus, Toronto, ON M5G 0C6, Canada*

[5]*Department of Computer Science, University of Toronto, 40 St George St., Toronto, ON M5S 2E4, Canada*

[6]*Department of Chemical Engineering & Applied Chemistry, University of Toronto, 200 College St., Toronto, ON M5S 3E5, Canada*

[7]*Institute of Medical Science, 1 King's College Circle, Medical Sciences Building, Room 2374, Toronto, ON M5S 1A8, Canada*

[8]*Canadian Institute for Advanced Research (CIFAR), 661 University Ave., Toronto, ON M5G 1M1, Canada*

[9]*NVIDIA, 431 King St W #6th, Toronto, ON M5V 1K4, Canada*

[†]*These authors contributed equally to this work.*

*Corresponding author email: yangb.bai@utoronto.ca

**Abstract**

Closed-loop experimentation requires more than selecting the next experiment. It must also determine which existing measurements and model-derived relationships are sufficiently supported to influence that decision. Here we introduce SPARC (Sparse Predictive Aggregation of Regression Components), a framework that continually revises this support as experimental evidence accumulates. Competing model families are evaluated through out-of-fold predictions, so that each model is tested on observations not used in its corresponding fit. Sparse posterior composition assigns model influence according to predictive contributions supported by the out-of-fold evidence, retaining complementary contributions from several model families. Candidate-dependent predictive dispersion and residual variability define trust, which modulates exploration alongside the predicted response. Independent diagnostics also prioritize existing measurements for prospective verification, with model-derived trust reported alongside them. Across eight published materials datasets, SPARC achieved higher $R^2$ than the corresponding literature-reported values in six, with relative gains of up to 77% without the density-functional-theory- or experiment-derived descriptors used by several published pipelines. In an automated alkaline oxygen-evolution-reaction campaign, 8 of 11 SPARC-selected experiments reached overpotentials below 300 mV, with a best value of 242 mV, compared with 18 of 125 initial experiments. Measurements prioritized for verification before repetition subsequently deviated by 76 mV on average, compared with 27 mV for matched controls. SPARC therefore turns experimental evidence into a basis for revising model influence, directing exploration and identifying measurements that warrant verification.

## Introduction

Machine learning is increasingly used to decide which experiments are performed next in scientific discovery.[1–9] Most closed-loop methods update their predictions as new measurements accumulate, but the representations, model families and initial assumptions about where to explore are often specified before an optimization campaign begins.[10–13] This assumption becomes restrictive in sparse experimental regimes, where even a small number of new observations can substantially change which relationships between experimental variables and performance remain supported by the data.[14–16] As these relationships change, so can the regions of experimental space that are worth exploring. Extending the search beyond the range represented by existing measurements introduces a further challenge, because experimental support weakens as the search moves away from what has already been observed. Closed-loop learning therefore requires more than sequential prediction updates. The modelling assumptions that shape a decision, and the support for extending exploration beyond the current experimental record, should also evolve as evidence accumulates.

Bayesian optimization (BO) provides a principled framework for sequential experiment selection by updating a probabilistic surrogate and balancing predicted performance against uncertainty.[13,17] However, its performance and acquisition behaviour depend on the representation and surrogate model chosen for the task.[14,16,18–20] Recent work has relaxed different parts of this constraint. Data-centric studies of out-of-distribution prediction show that generalization depends strongly on the diversity and coverage of experimental data, and that uncertainty can guide acquisition towards less-explored chemical space.[21] Recent probabilistic language-model approaches further allow the representation itself to adapt during optimization, using experimental outcomes and uncertainty to reorganize the latent space for more sample-efficient search.[22] Furthermore, the representation of the language model itself can be used as the representation directly.[23] These advances make experimental search more adaptive. A complementary challenge remains because the influence of alternative models must also be reconsidered as new evidence arrives. When a particular representation and surrogate family are chosen at the outset, their predictions and uncertainties shape where subsequent experiments are acquired. This feedback can progressively concentrate exploration in regions the current model favours and reinforce its underlying assumptions. Other models that remain plausible under the same sparse data may therefore receive less opportunity to be tested, even when they imply different relationships elsewhere in the design space. Reliable closed-loop learning should instead allow new experimental evidence to revise which models remain influential and, through their local agreement, which regions of the search space remain sufficiently supported for further exploration.

In this work, we introduce SPARC (Sparse Predictive Aggregation of Regression Components), which incorporates this evolving model support into the experimental loop. SPARC evaluates competing model families using out-of-fold predictions: each observed sample is predicted by a model trained without that sample, providing a direct test of how well each model family predicts data not used in its fitting. Sparse posterior composition assigns relative influence to the candidate models according to their joint predictive contributions. Several families can remain influential when their complementary predictions are supported by the data. SPARC then evaluates the resulting predictive dispersion at each candidate. The retained models can agree in one region and diverge in another, making uncertainty in their composed prediction depend on location. Candidate-dependent dispersion and a residual scale from prediction errors define candidate-specific trust. As additional experimental data enter the loop, predictive support and candidate-specific trust estimates are updated, allowing the basis for subsequent decisions to change with the accumulated evidence. Trust modulates uncertainty-driven acquisition as the search enters less-supported regions. In parallel, independent diagnostics prioritize existing measurements for prospective verification. SPARC thus connects empirical model support to experimental priorities while directing additional scrutiny towards the existing record.

We evaluate SPARC across eight published materials datasets and in sequential, extrapolative and automated experimental settings. Across molecular, polymeric and inorganic systems, SPARC achieves strong predictive performance without the additional density-functional-theory- or experimentally derived descriptors used by several published pipelines. In sequential search and controlled extrapolation, SPARC continues to prioritize high-performing candidates while becoming more uncertain as the search moves beyond regions supported by the initial experiments. We then test these decisions prospectively in an automated alkaline oxygen-evolution-reaction campaign. Eight of eleven SPARC-selected experiments achieve overpotentials below 300 mV at 10 mA $cm^{-2}$, compared with 18 of 125 experiments in the initial dataset, with the lowest overpotential reaching 242 mV. Measurements prospectively prioritized for repetition subsequently showed a higher mean absolute original–repeat difference than matched controls (76 versus 27 mV), driven by three deviations above 130 mV. Together, these results establish the value of evidence-controlled modelling for prediction, experimental prioritization and prospective measurement verification.

## Results

### Evidence-guided model composition

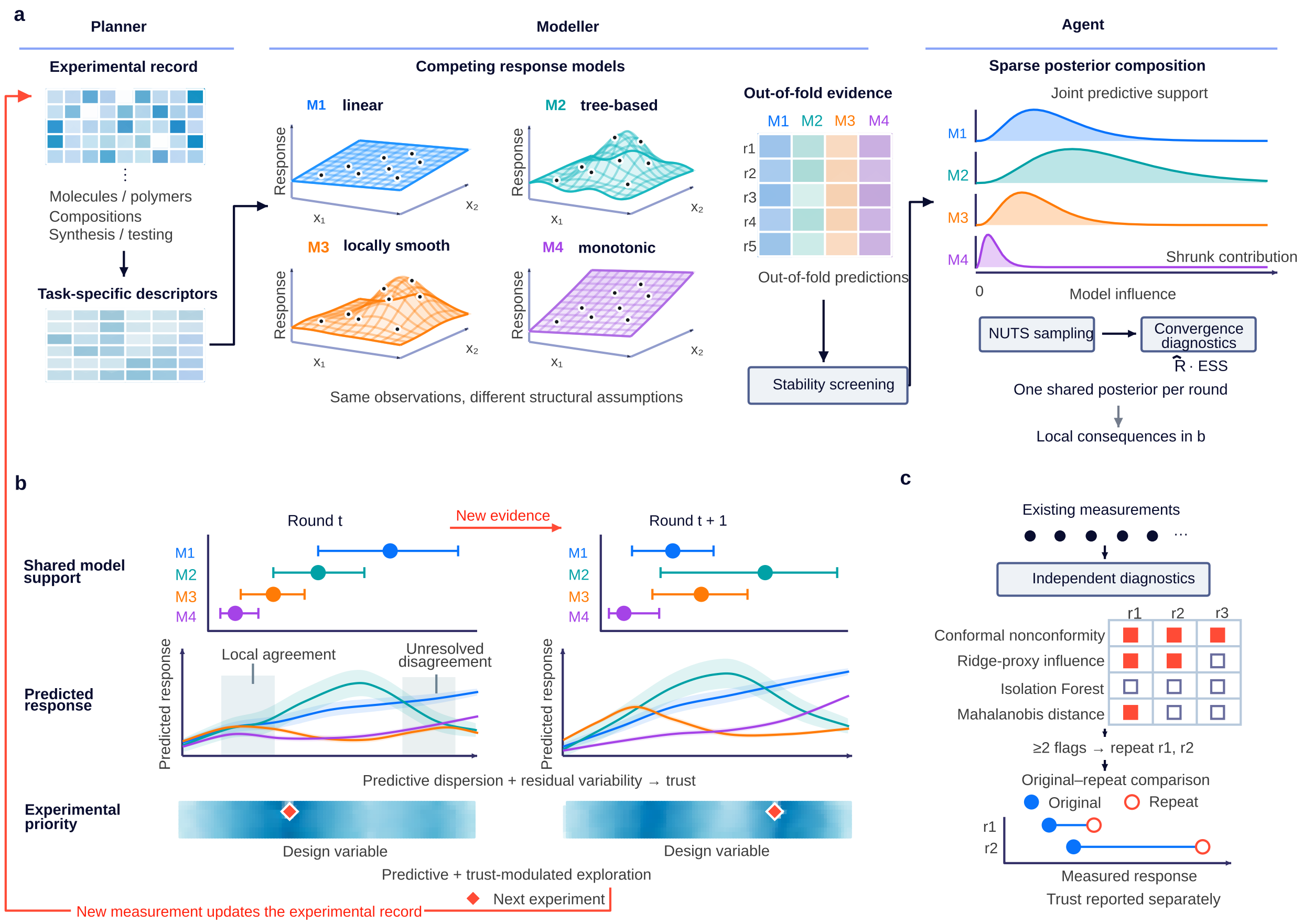

**Fig. 1 | SPARC makes model influence and experimental decisions responsive to evidence. a,** Out-of-fold evidence from a fixed model library determines the sparse posterior composition in the Bayesian benchmark core. Weights are shared across candidates within a training state and can be revised when the experimental record changes. b**,** Retained response models can agree locally and diverge elsewhere. Candidate-dependent predictive dispersion and a residual scale define trust, which modulates exploration alongside the predicted response. Sequential and laboratory analyses use heterogeneous-bootstrap predictive statistics; controlled extrapolation uses three Gaussian-process experts weighted by cross-validated predictive likelihoods. **c,** Independent diagnostics prioritize existing measurements for prospective repetition, with trust reported separately. All curves, point positions and weight bars are schematic.

SPARC makes the influence of competing models responsive to out-of-sample experimental evidence (Fig. 1). Its Planner–Modeller–Agent workflow separates representation, predictive evaluation and decision-making within a single pipeline. The Planner constructs task-specific representations, the Modeller evaluates competing response-surface models, and the Agent combines their predictions and quantifies support for subsequent actions. Molecular, compositional and process descriptors accommodate the different materials tasks. The shared principle is to assess what each model contributes on the observed task before using the resulting prediction to guide an experiment.

The Bayesian composition evaluates linear, tree-based nonlinear, locally smooth and monotonic response models against a common out-of-fold evidence matrix. Each entry records a prediction for an observation excluded from the corresponding model fit. Stability screening identifies a reproducible active set, and non-negative posterior composition assigns relative influence within that set. The composition uses hierarchical half-Cauchy scale priors, a construction also used in high-dimensional Bayesian optimization.[24] Complementary models can remain influential even when their response surfaces differ. During benchmarking, an outer leave-one-out evaluation repeats preprocessing, model selection and posterior fitting within each training partition. The held-out observation therefore evaluates the entire composition procedure.

Model support determines which predictors contribute, whereas candidate-specific trust determines how their uncertainty enters an experimental decision. Within the Bayesian composition, all candidates share the same posterior distribution over model weights at a given training state. Propagating posterior weight draws through the retained response surfaces produces candidate-dependent composition uncertainty. SPARC combines this dispersion with a robust residual scale to define trust. The residual term retains information about variation unexplained by the available models and descriptors. Trust modulates the exploration contribution alongside the predicted response, thereby changing experimental priorities within the available candidate domain. Numerical definitions and the distinct uncertainty estimates used by each implementation are given in Methods.

We evaluated complementary functions of this workflow across prediction, acquisition and verification. Predictive benchmarks tested the full Bayesian posterior composition. Retrospective finite-pool search and laboratory recommendations tested trust-modulated acquisition with a heterogeneous bootstrap ensemble. Controlled extrapolation used three Gaussian-process experts whose weights were updated using cross-validated predictive likelihoods. The acquisition implementations each supplied a predictive mean and dispersion to the decision rule. Separate feature-strategy comparisons assessed representation choices, and independent diagnostics prioritized existing measurements for verification. Together, these studies evaluate how SPARC uses evidence to guide model composition, trust to modulate acquisition and independent diagnostics to prioritize measurements for prospective verification.

## Predictive benchmarking across materials datasets

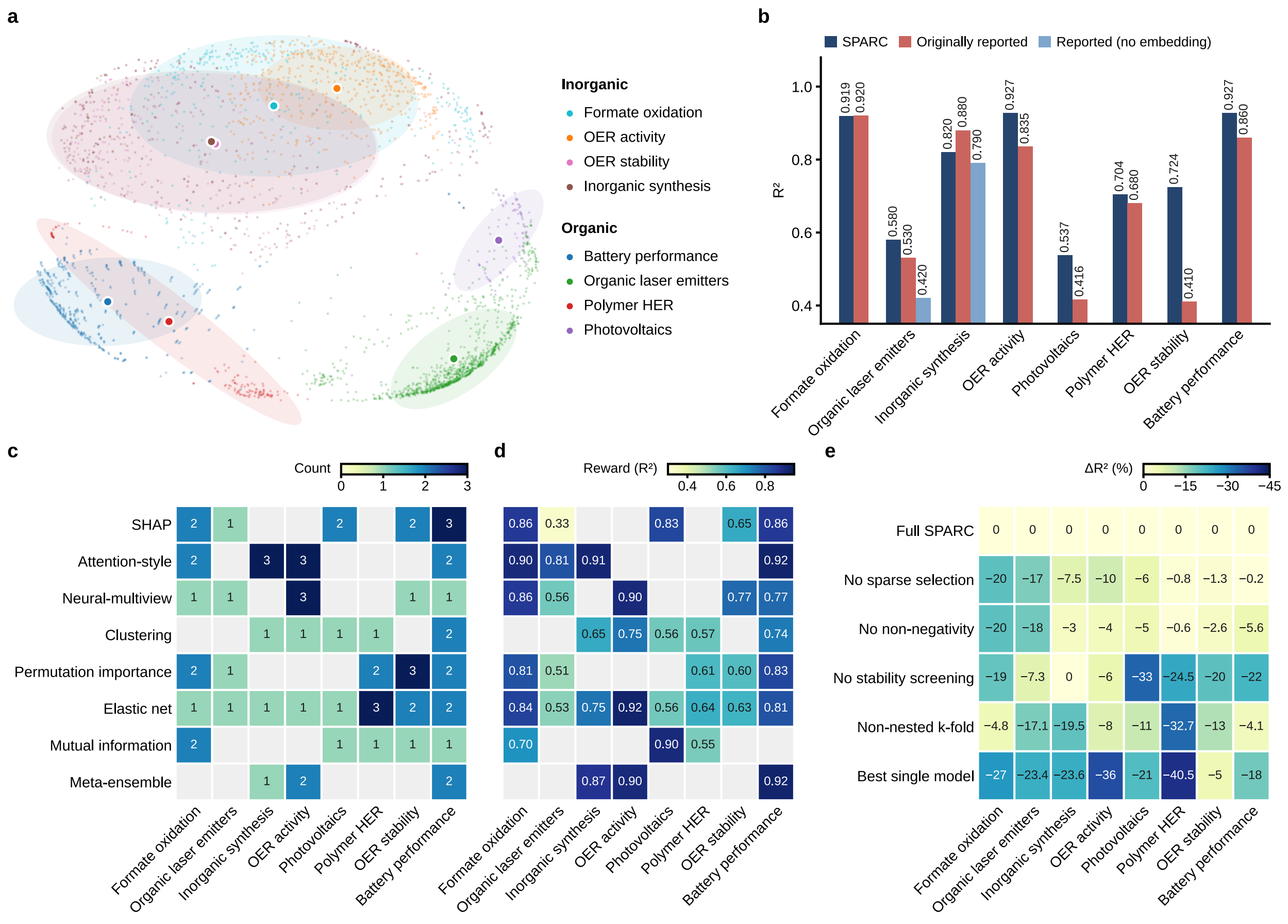

**Fig. 2 | Predictive performance, representation choices and the contribution of model composition. a,** Two-dimensional visualization of eight published datasets spanning molecular, polymeric and inorganic materials tasks. The panel illustrates benchmark heterogeneity; it was not used for model fitting, and between-dataset distances are not interpreted as a universal chemical metric. **b,** Predictive $R^2$ for SPARC and the corresponding literature-reported pipelines; for inorganic synthesis, a binary classification of gel formation in sol–gel synthesis, the values in b, d and e refer to classification accuracy. Literature values provide context because evaluation splits were not uniform across source studies. **c,** Selection counts for eight feature-fusion and feature-weighting strategies under nested evaluation. **d,** Mean recorded predictive reward ($R^2$) for each evaluated strategy–dataset combination; grey cells indicate combinations without a reported reward. **e,** Relative change in $R^2$ (%) from full SPARC for core-component ablations and simplified alternatives.

Eight published materials datasets provided a cross-domain test of evidence-controlled model composition, spanning formate oxidation,[25] organic laser emitters,[8] inorganic synthesis,[12] OER activity,[26] photovoltaics,[27] polymer HER,[28] OER stability[12] and battery performance.[29] Each task retained its domain-appropriate descriptors (Supplementary Note 1 and Supplementary Table 1). SPARC and the reimplemented comparison methods were evaluated on matched outer test observations under nested validation. Figure 2a summarizes the heterogeneous molecular, compositional and process inputs covered by these tasks.

SPARC reached predictive performance comparable to or exceeding the values reported for seven of the eight source pipelines (Fig. 2b and Supplementary Fig. 1). For OER stability, $R^2$ was 0.724 compared with the literature-reported 0.410, an absolute difference of 0.314 and a relative difference of approximately 77%. Across four tasks whose published pipelines used auxiliary density-functional-theory- or experimentally derived descriptors, relative gains ranged from 3.5% to 77% without those additional calculations or measurements. SPARC also exceeded the reported battery result obtained with a feature-engineered stacked ensemble. These literature values provide task-specific context under the source studies' evaluation protocols. Across these varied settings, SPARC delivered strong predictive performance using the task descriptors already available.

The feature-strategy evaluations showed that representation complexity alone did not determine predictive value (Fig. 2c,d). Across eight feature-fusion and feature-weighting strategies, no strategy was universally preferred. Meta-Ensemble achieved the highest mean reward among the recorded evaluations, whereas several simpler alternatives were selected more often. Selection frequency and mean recorded reward captured different aspects of strategy performance. Evaluating each treatment on the target task allowed simple strategies to remain competitive with more elaborate feature combinations.

Component comparisons showed that the contribution of model composition varied across tasks (Fig. 2e). Removing sparse selection or non-negativity reduced $R^2$ on all eight tasks, with relative losses of 0.2–20.0% and 0.6–20.0%, respectively. Removing stability screening reduced $R^2$ on seven tasks, with losses of 6.0–33.0%, and produced no change at the reported precision on inorganic synthesis. Selecting only the best single model reduced $R^2$ on all eight tasks, with losses of 5.0–40.5%. The tested non-nested k-fold variant also yielded lower $R^2$ on all eight tasks, with relative losses of 4.1–32.7%. These comparisons support the predictive value of model composition and its component constraints, while showing that the magnitude of their contributions depends on the task.

## Predictive value of additional information

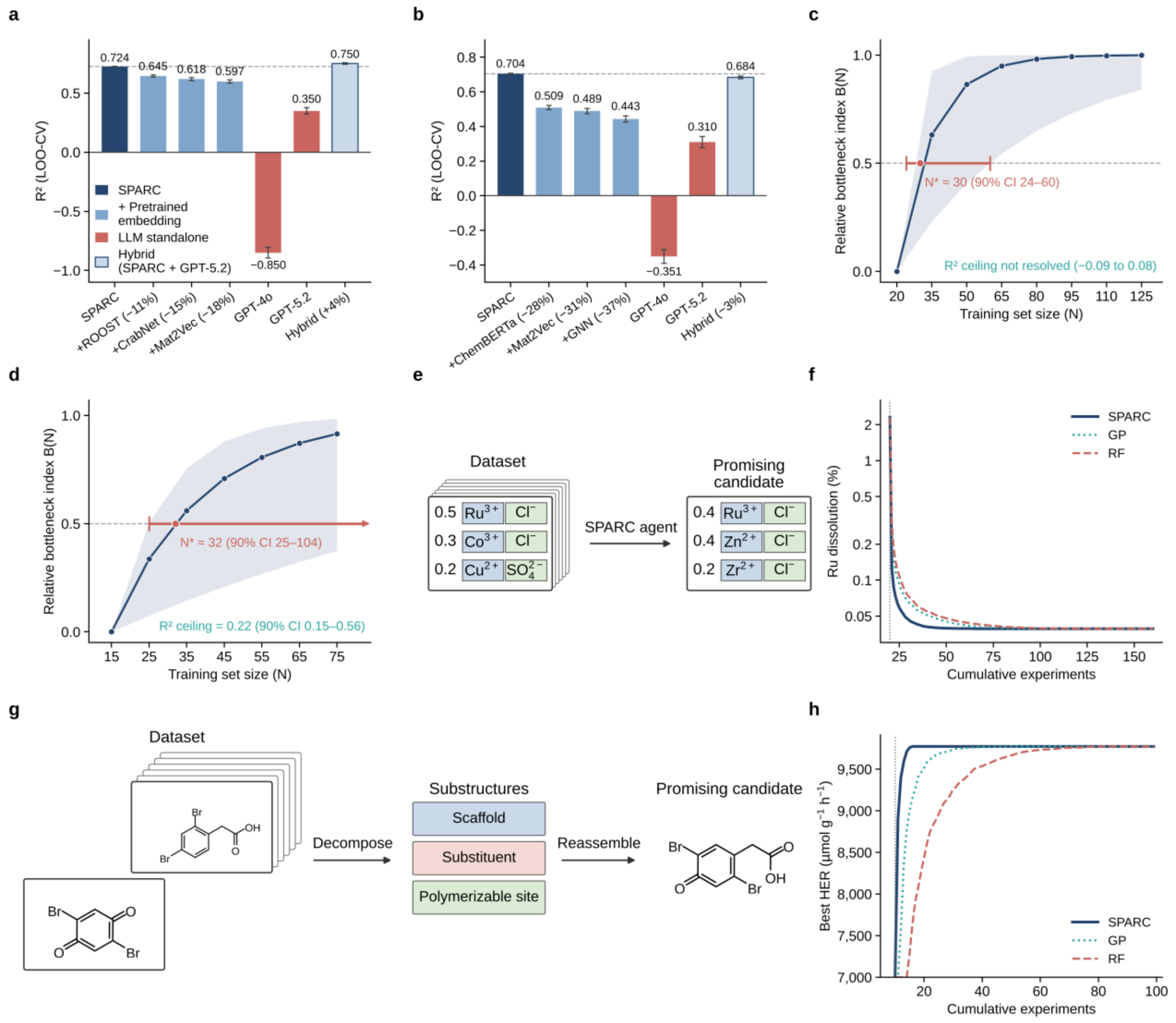

**Fig. 3 | Evidence-guided evaluation of representations, data and candidate acquisition. a,b,** Leave-one-out cross-validated $R^2$ for alloy-oxide OER stability (a) and polymer HER (b), comparing SPARC with pretrained-representation augmentation, standalone LLM approaches and hybrid LLM–SPARC variants. Percentages show change relative to SPARC; error bars denote ±1 s.d. **c,d,** Relative bottleneck index B(N) for OER stability (c) and polymer HER (d), obtained from repeated held-out learning curves for a fixed three-model nonlinear ensemble. Shading denotes 90% bootstrap intervals and markers indicate fitted transition sizes N*. The model configuration and evaluation protocol differ from those in a,b. **e,g,** Candidate spaces constructed from precursor–composition recipes (e) and compatible polymer fragments (g). **f,h,** Retrospective best-so-far trajectories in finite labelled pools for Ru dissolution (f; lower is better) and polymer HER (h; higher is better), comparing the SPARC bootstrap implementation, Gaussian-process optimization and random forest. All strategies started from the same 20 Ru-dissolution experiments or 10 polymer candidates; evaluation counts include initialization. Reported discovery counts denote the first evaluation within 0.1% of the pool optimum.

We next tested whether pretrained representations added predictive value beyond SPARC's task-specific descriptors for alloy-oxide OER stability and polymer HER (Fig. 3a,b and Supplementary Note 7). We separately augmented the OER-stability descriptors with ROOST,[30] CrabNet[31] or Mat2Vec[32] representations, and the polymer descriptors with ChemBERTa,[33] Mat2Vec or a graph-convolutional-network representation.[34] SPARC's task-specific descriptor configuration outperformed every tested augmentation. Relative $R^2$ decreased by 11–18% for OER stability and 28–37% for polymer HER after augmentation. The original descriptors therefore supplied a stronger predictive basis under the evaluated training conditions.

SPARC also outperformed the standalone GPT-based approaches on both tasks (Supplementary Figs 2 and 3 and Supplementary Table 6). In the hybrid setting, GPT-5.2 selected among SPARC's existing feature strategies and model hypotheses. This increased OER-stability $R^2$ from 0.724 to 0.750 but reduced polymer-HER $R^2$ from 0.704 to 0.684. Language-model guidance thus added value for OER stability, whereas the original SPARC configuration remained stronger for polymer HER. The contrast reinforces the value of testing an additional decision layer against the task-specific baseline.

We examined the return from additional training records using fixed descriptors and a three-model ensemble of Extra Trees, random forest and gradient boosting (Fig. 3c,d and Supplementary Note 8 and Supplementary Table 7). Repeated held-out learning curves were summarized by the relative bottleneck index B(N), the fraction of fitted improvement attained between the smallest training set and the asymptote. This diagnostic isolates data-volume changes within an unchanged predictive pipeline. Its model configuration and evaluation protocol differ from the full SPARC comparisons in Fig. 3a,b.

Both learning curves showed diminishing returns within the evaluated data regime. The OER-stability fitted $R^2$ ceiling was not resolved above zero, whereas polymer HER retained a positive but limited fitted ceiling. At the largest evaluated polymer training size, ten additional records provided a fitted gain of approximately 0.008 in $R^2$. These estimates characterize saturation of the tested fixed pipeline rather than a ceiling on SPARC. Together with the representation comparisons, they show why the predictive value of an information source must be assessed on the task itself.

## Trust-modulated sequential optimization

We next tested how effectively predictive information was translated into experimental acquisition order. Retrospective searches used finite labelled pools for alloy-oxide OER stability (Ru dissolution) and polymer HER (Fig. 3e–h and Supplementary Note 9). Candidate descriptions captured precursor identities and composition patterns for alloy oxides, and scaffold, substituent and polymerizable-site information for polymers. Archived outcomes were concealed and revealed only after selection, allowing sequential decisions to be compared within the same candidate pool.

SPARC's bootstrap implementation was compared with Gaussian-process optimization and random forest from identical initial sets of 20 Ru-dissolution experiments or 10 polymer candidates. At each acquisition step, a candidate was selected, its concealed outcome was revealed and the model was updated. Discovery was defined as the first evaluation at which the best-so-far response was within 0.1% of the pool optimum, with cumulative counts including initialization.

SPARC reached the near-optimality criterion after 66 evaluations for Ru dissolution, compared with 113 for Gaussian-process optimization and 126 for random forest (Fig. 3f). For polymer HER, the corresponding counts were 16, 39 and 76 (Fig. 3h). These counts correspond to approximately 42–48% fewer evaluations for Ru dissolution and 59–79% fewer for polymer HER relative to the two comparators. All counts include the shared initialization. Each count comes from a single search started from one initial set per task, so these comparisons are case studies; estimating average search efficiency would require repeated initializations (Supplementary Note 9). In both reported trajectories, SPARC brought near-optimal candidates forward in the acquisition order, converting the available records into a lower experimental search budget.

## Candidate prioritization under controlled extrapolation

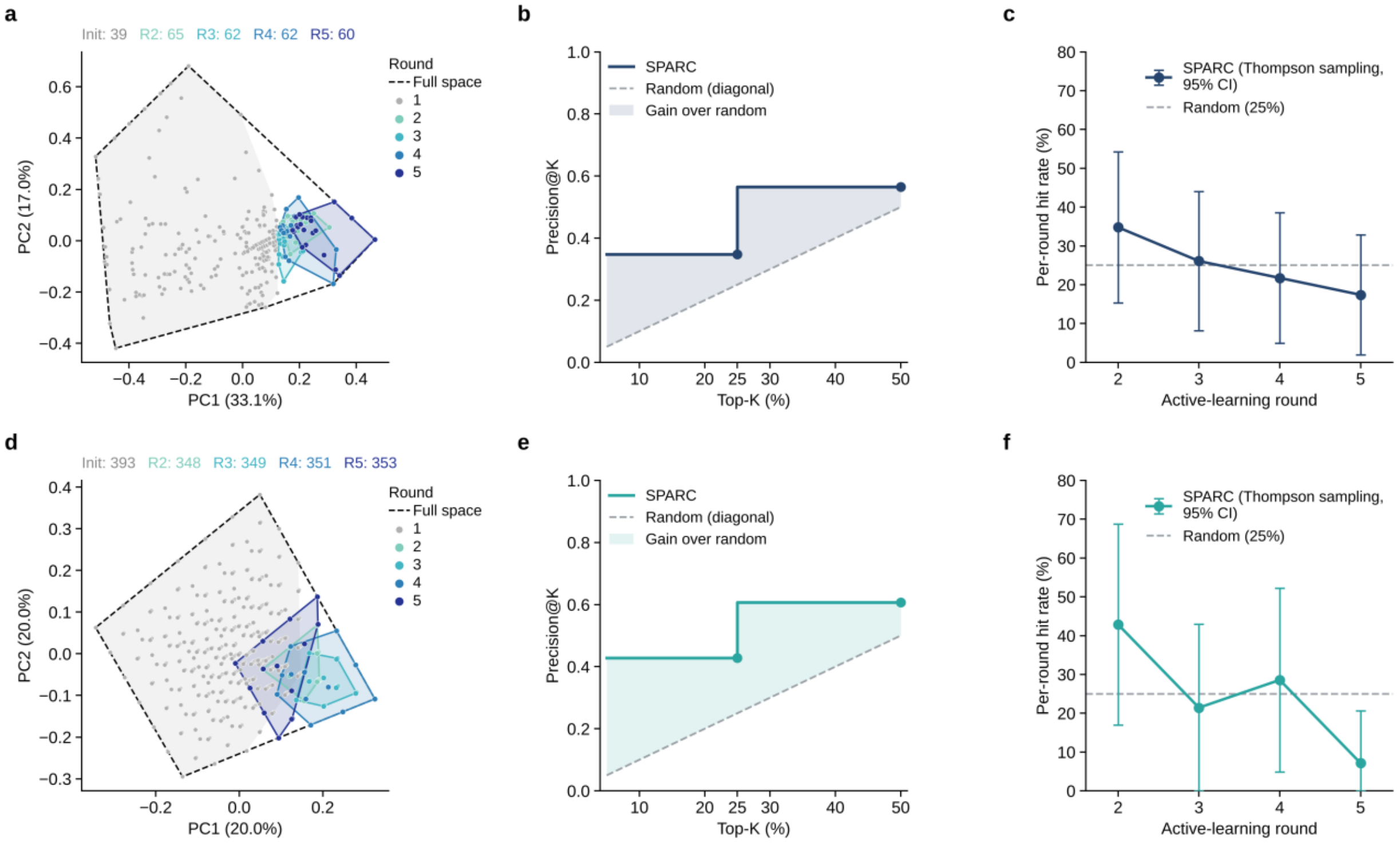

**Fig. 4 | SPARC retains informative candidate rankings under controlled extrapolation. a,d,** Principal-component projections of the formate-oxidation (a) and OER-activity (d) design spaces. Grey points mark the initial region: candidates up to the 70th percentile of the first principal component of the inputs (formate oxidation) or at cobalt levels 1–3 (OER activity). Coloured points and convex hulls show sequentially selected batches, and dashed outlines show the complete candidate space. Values above each panel report the mean measured target of the initial set and each acquired batch. **b,e,** Precision@K at the first two complete-batch cutoffs, corresponding to acquisition of 25% and 50% of the extrapolation pool. Step curves summarize these batch-level evaluations and do not imply within-batch ranking. **c,f,** Per-round hit rate, defined as the fraction of selected candidates in the true top quartile of the original pool. Error bars denote ±1.96 binomial standard errors; dashed lines mark the 25% random expectation. The three-expert Gaussian-process implementation selected four batches of 23 formate-oxidation candidates or 14 OER-activity candidates. Batches were fixed before labels were revealed and models were updated between rounds. The fourth acquisition batch (round 5 in c,f) exhausted each pool.

We then challenged the workflow to prioritize candidates beyond the operating window represented by the initial experiments. Controlled extrapolation tasks were constructed for formate oxidation and OER activity (Fig. 4 and Supplementary Note 10). The splits were defined from the inputs only. For formate oxidation, candidates above the 70th percentile of the first principal component of the inputs formed the extrapolation pool; for OER activity, compositions at cobalt levels 4–7 formed the extrapolation pool and levels 1–3 the initial region. This yielded 214 initial and 92 extrapolation candidates for formate oxidation, and 406 initial and 56 extrapolation candidates for OER activity. The task therefore tested expansion along the existing experimental variables.

The three-expert Gaussian-process implementation selected four batches of 23 formate-oxidation candidates or 14 OER-activity candidates using trust-modulated Thompson sampling. Each entire batch was fixed before any of its labels were revealed, after which the new observations were used to update the experts and their weights. Candidate ranking was consequently evaluated at complete-batch cutoffs rather than inferred within a batch.

The workflow retained informative candidate rankings as predictive uncertainty increased beyond the initial region. Extrapolation-to-initial-region uncertainty ratios, computed with the initial-region predictive standard deviation at the training points as the denominator, were 1.27 for formate oxidation and 1.75 for OER activity (Supplementary Table 8 and Supplementary Figs 5 and 6). The first acquired batch, covering one-quarter of each extrapolation pool, contained 35% and 43% true top-quartile candidates, respectively, against the 25% random expectation (Fig. 4b,e). The three-expert implementation thus combined increased uncertainty with early enrichment for high-performing conditions. Because the fourth batch exhausted each pool, the decision benefit was earlier access to these conditions, as resolved at complete-batch cutoffs (Fig. 4c,f). Hit rates in later rounds therefore fell below 25% as the remaining top-quartile candidates were depleted without replacement.

## Prospective validation in an automated OER campaign

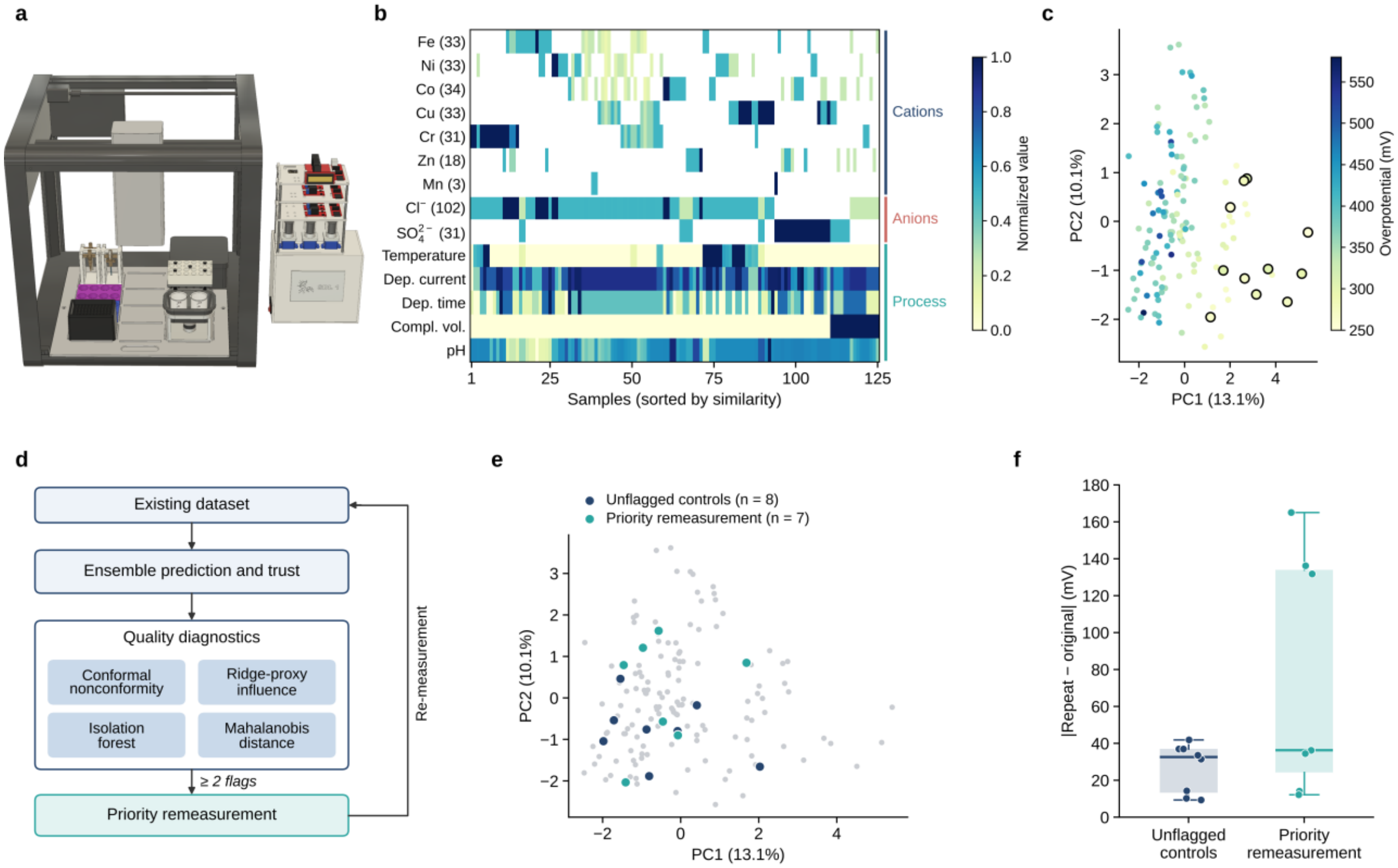

**Fig. 5 | Prospective catalyst selection and measurement verification in an automated OER campaign.** **a,** Automated electrodeposition and electrochemical characterization platform. **b,** Initial dataset of 125 experiments spanning seven transition-metal cations, two anion chemistries and five synthesis variables. **c,** Principal-component projection coloured by overpotential at 10 mA $cm^{-2}$. Black-edged circles mark 11 recommendations fixed before measurement; eight produced overpotentials below 300 mV and the best reached 242 mV. **d,** Verification workflow reporting candidate-specific trust alongside four diagnostics: conformal nonconformity, Ridge-proxy influence, Isolation Forest score and Mahalanobis distance, which measures covariance-adjusted deviation from the feature-space centre to identify multivariate outliers. Measurements with at least two diagnostic flags were prioritized for repetition; trust was reported separately. The return arrow denotes potential reuse after verification; repeat measurements were evaluated independently and were not used in a subsequent acquisition round here. **e,** Locations of unflagged controls and priority-remeasurement samples. **f,** Absolute differences between original and repeated overpotentials for eight matched controls and seven priority samples. Group assignments were fixed before the repeat outcomes were measured.

An automated multielement alkaline OER campaign tested two prospective decisions: which new conditions to measure and which existing records to repeat. The platform integrated electrodeposition with electrochemical characterization,[35] and the initial record comprised 125 independent experiments spanning seven transition-metal cations, two anion chemistries and five synthesis variables. These variables were deposition temperature, current, time, complexing-agent volume and precursor pH (Fig. 5a,b). Recommendations and verification-group assignments were fixed before the respective measurement outcomes became available.

SPARC's bootstrap implementation recommended 11 new conditions before their outcomes were measured. Eight recommendations (72.7%) achieved overpotentials below 300 mV at 10 mA $cm^{-2}$, compared with 18 of 125 initial experiments (14.4%). The best condition reached 242 mV (Fig. 5c, Supplementary Note 11 and Supplementary Fig. 7). Relative to the historical record, the recommended batch delivered approximately fivefold enrichment in conditions meeting the activity threshold. This prospective result demonstrates productive allocation of the next experimental batch using the existing record.

Independent diagnostics directed verification towards existing measurements before their repeatability was known (Fig. 5d, Supplementary Note 12 and Supplementary Fig. 8). Conformal nonconformity, Ridge-proxy influence, Isolation Forest anomaly scores and Mahalanobis distance supplied four complementary flags. Measurements with at least two flags formed the priority-remeasurement group. Controls had no flags and were matched in feature space. Candidate-specific trust was reported alongside the diagnostics as a separate quantity. All group assignments preceded repetition, providing a prospective test of verification priority.

The priority group captured the largest original–repeat differences (Fig. 5e,f). All eight controls differed from their original values by less than 42 mV, with a mean absolute difference of 27 mV. The seven priority measurements had a mean absolute difference of 76 mV, including three deviations above 130 mV (Fig. 5f). Median absolute differences were closer, at 32.5 mV for controls and 36.3 mV for priority measurements. The value of the screening was therefore concentrated in identifying records with large observed deviations. These outcomes demonstrate that prospective diagnostic prioritization can direct scrutiny towards the large-deviation tail of the experimental record.

Together, the campaign demonstrated two complementary allocations of experimental effort: finding active conditions and verifying selected records in the existing dataset. Both decisions were fixed before the measurements that evaluated them. The acquisition and verification tests were evaluated separately; repeat measurements were not fed into a subsequent acquisition round.

## Discussion

In this work, we address a limitation of closed-loop learning in sparse experimental regimes. New measurements routinely update predictions, but the modelling assumptions that shape subsequent decisions are often chosen before the campaign begins. SPARC instead allows the influence of competing model families to change as experimental evidence accumulates. Each model family is evaluated by its ability to predict observations that were not used for fitting, so its contribution reflects out-of-sample evidence rather than training performance alone. The retained models can nevertheless agree in some regions of the experimental space and diverge in others. SPARC combines candidate-dependent predictive dispersion with residual variability to define trust and modulate exploration. This separates support for a model composition from the strength of the prediction at a particular candidate.

The results show that this distinction remains useful across chemically diverse tasks and changing experimental regimes. SPARC maintains strong predictive performance across molecular, polymeric and inorganic datasets without requiring several density-functional-theory- or experimentally derived descriptors used in previous pipelines. More complex representations, including the tested pretrained and language-model-based alternatives, do not provide a consistent advantage. The fixed-pipeline learning curves also show diminishing returns from adding records of the same type. The acquisition studies extend this evaluation to experimental decisions: SPARC reaches high-performing candidates efficiently and becomes more uncertain as controlled extrapolation moves beyond the region represented by the initial experiments, while retaining useful ranking of promising candidates. The prospective alkaline oxygen-evolution-reaction campaign demonstrates productive candidate selection in the laboratory, with 8 of 11 recommended experiments reaching overpotentials below 300 mV at 10 mA $cm^{-2}$ and a best value of 242 mV. Measurements identified before repetition for priority remeasurement subsequently showed a higher mean absolute original–repeat difference than matched controls (76 versus 27 mV), driven by three deviations above 130 mV.

Several limitations define the current scope of SPARC. The framework operates over researcher-specified representations and model-family libraries and therefore revises the contribution of available models rather than generating fundamentally new modelling hypotheses during a campaign. Candidate-specific trust is task-dependent and should not be interpreted as a calibrated probability of prediction correctness or measurement error. The controlled-extrapolation experiments probe expansion beyond the initial ranges of the same experimental variables rather than transfer across unrelated chemistries, and broader distribution shifts will require explicit treatment of changes in chemistry, experimental protocol and laboratory conditions. Scaling to substantially larger datasets and model libraries will also require more efficient estimation of out-of-sample evidence. Measurements prioritized for repetition should likewise be interpreted as requiring additional evidence rather than as observations already shown to be incorrect.

More broadly, these findings support a view of closed-loop science in which the model underlying an experimental decision remains open to revision throughout the campaign. In sparse-data regimes, disagreement among models that remain empirically viable records relationships that the available experiments have not yet resolved. Model disagreement provides a measurable signal of unresolved predictive relationships, while prospective verification directs experiments towards existing records that require further scrutiny. SPARC combines these functions in a practical workflow for revising model influence and experimental priorities as evidence accumulates. This evidence-adaptive modelling layer could integrate with experiment orchestration agents such as La Agente Óptima,[9] enabling future systems to revise not only models and representations but also the experimental design strategy itself as a campaign evolves. This points towards self-driving laboratories in which accumulating evidence continuously refines both scientific predictions and the decision process built around them. What ultimately closes the gap is the generation of truly novel hypotheses. When a system that generates hypotheses is connected to a combined system exploiting SPARC and La Agente Óptima, we could enter the next step of increased autonomous discovery.

## Methods

### Study design and datasets

We evaluated SPARC on eight published materials datasets, two retrospective finite-pool optimization tasks and controlled extrapolation on two datasets. A prospective automated oxygen evolution reaction (OER) campaign evaluated candidate selection and independent measurement verification. The predictive benchmarks and pretrained-representation comparisons used full Bayesian posterior composition under nested outer leave-one-out validation. The finite-pool studies and OER campaign used a heterogeneous bootstrap ensemble for computationally tractable acquisition. Controlled extrapolation used three Gaussian-process experts weighted by cross-validated predictive likelihoods. These implementations evaluated complementary functions: predictive composition, sequential acquisition and exploration beyond the initial operating window. Candidate-specific trust combined each implementation's predictive dispersion with a residual scale and modulated acquisition as defined below. Feature-strategy adaptation was evaluated separately in Fig. 2c,d. Main-figure analyses did not use training-time soft gating of potentially unreliable observations.

The eight benchmark datasets covered formate oxidation, OER activity, OER stability, inorganic synthesis, battery performance, organic laser emitters, polymer hydrogen evolution reaction (HER) and photovoltaic materials. They spanned molecular, polymeric and inorganic systems with different sample sizes, target scales, experimental sources and descriptor sets. Molecular inputs comprised fingerprints and physicochemical descriptors; inorganic inputs comprised composition-derived elemental descriptors; synthesis and testing conditions were encoded as process variables. We fitted all preprocessing transformations and feature filters using only the corresponding training partition. These steps included missing-value imputation, categorical encoding, feature scaling and removal of zero-variance features.

For predictive benchmarking, SPARC and all reimplemented baselines were evaluated on identical outer test observations under the nested leave-one-out protocol described below. Values taken directly from source publications were reported only as literature references, not as matched-split baselines. Unless stated otherwise, we computed $R^2$ from the assembled outer predictions and corresponding held-out responses, rather than averaging scores over single-observation folds. The learning-curve and acquisition studies followed the separate protocols described below. Dataset descriptions, target definitions and implementation details are provided in the Supplementary Information (Supplementary Table 2).

The two-dimensional map in Fig. 2a visualized benchmark heterogeneity and was not used for model fitting. Distances in this map were not interpreted as universal chemical distances across molecular, polymeric and inorganic systems.

**Evidence-controlled posterior composition**

SPARC represented alternative response-surface structures through 12 candidate models grouped into four structural families (Supplementary Table 4). Posterior weights were assigned to individual candidate models. Linear models comprised Ridge, Lasso, elastic-net, Bayesian-ridge and Huber regressions. Tree-based nonlinear models comprised random forest, Extra Trees and gradient boosting. Locally smooth models comprised radial-basis-function support-vector regression, distance-weighted k-nearest-neighbour regression and a Matérn-5/2 Gaussian process. A monotonic gradient-boosting regressor represented the fourth structural class.

For each candidate model m, support was estimated from leave-one-out predictions. We fitted the model without labelled observation i and predicted its omitted response:

$$P_{im} = f_m^{-i}(x_i) \quad (1)$$

The n × M matrix P collected labelled observations as rows and candidate models as columns. Thus, entry $P_{im}$ was always generated by a model fit that did not use response $y_i$.

Predictive performance was estimated by nested outer leave-one-out validation. For each outer test observation, we repeated every data-dependent operation using only the remaining observations. These included preprocessing, feature-strategy selection when enabled, hyperparameter selection, stability screening, evidence-matrix construction and posterior-weight estimation. For predictions from the Bayesian composition at a given training state, we constructed an internal leave-one-out evidence matrix from all currently labelled data. After estimating the posterior composition, we refitted each retained model to the complete labelled set before predicting unmeasured candidates.

Before posterior composition, we assessed model-selection stability using 50 subsamples drawn without replacement, each containing 80% of the available training records.[36] For each subsample, we constructed out-of-fold evidence and used an $\ell_1$-regularized non-negative fit to identify the support set. Candidate models selected in at least 60% of the subsamples entered the active set. Stability screening was omitted when fewer than 20 training records were available. If no model met the 60% criterion, we retained the five models with the highest selection frequencies. SPARC then allocated non-negative simplex weights across the active models using normalized magnitudes of hierarchical scale-mixture coefficients.

$$\alpha_m = |a_m|, w_m = \frac{\alpha_m}{\sum_j \alpha_j}; y|P, w, \sigma^2 \sim \mathcal{N}(Pw, \sigma^2 I) \quad (2)$$

For each retained model, we assigned $a_m \sim \mathcal{N}(0, {s_m}^2)$ , $\rho \sim \textit{Half-Cauchy}(0,1)$ , $s_m \sim \textit{Half-Cauchy}(0, \rho c_m)$ and $\sigma^2 \sim \textit{Inverse-Gamma}(1,1)$. The positive family-scale parameter $c_m$ was specified by structural family, with models in the same family sharing the assigned value. It entered the local-scale prior directly and defaulted to 1. Normalization converts coefficient magnitudes into relative contributions that sum to one. The common multiplicative scale cancels, leaving model-specific scale variation to determine the induced prior over relative weights. We sampled the posterior with the No-U-Turn sampler (NUTS), using four chains, 1,000 warm-up iterations and 2,000 retained draws per chain. We assessed sampling convergence for each normalized model weight using the criteria $\hat{R} < 1.01$ and an effective sample size above 400. If either criterion failed, we repeated sampling once with 2,000 warm-up iterations. Predictions used posterior-mean weights and retained posterior draws quantified uncertainty in the model composition.

Exact exclusion of models is imposed by the preceding support-set screening. Within the active set, the continuous prior assigns positive weights almost surely. The shared multiplicative scale cancels under normalization, so relative family-scale values, rather than the shared scale alone, determine family-specific prior preferences.

**Trust and feature-strategy adaptation**

Each training state defined one posterior distribution over model weights, shared by all candidate inputs. Propagating its draws through the retained response surfaces produced a distribution of composed predictions at each candidate. Candidate-level variation therefore depended jointly on differences among model predictions and uncertainty in their relative weights. Several models could contribute without requiring their response surfaces to coincide.

For fixed fitted response surfaces, composition variance is

$$ {u_{ep}}^2(x) = f(x)^T Cov(w|P, y) f(x) \quad (3) $$

where f(x) is the vector of retained model predictions. This quantity measures uncertainty in the composed mean conditional on the fitted models; it is not the weighted variance of the individual model predictions and does not include all within-model parameter uncertainty.

SPARC combined a candidate-dependent dispersion term with a residual scale to quantify support for each candidate prediction. Under the Bayesian composition, each posterior weight draw produced $\mu^{(s)}(x) = \sum_m w_m^{(s)} f_m(x)$ , and $u_{ep}(x)$ was the standard deviation across draws, quantifying posterior composition uncertainty. This differs from the spread of model predictions at a fixed weight vector. Finite-pool and laboratory acquisition used bootstrap predictive standard deviations. Controlled extrapolation used Gaussian-process mixture standard deviations, including within-expert variance and between-expert disagreement. The residual scale was $u_{res} = 1.4826 \, median_i |e_i - median(e)|$ . For Bayesian composition, residuals used a truncated importance-sampling approximation to leave-one-out prediction at the composition layer,

conditional on the precomputed evidence matrix. The outer leave-one-out evaluation separately refitted the complete predictive pipeline for each test observation. For the bootstrap ensemble, residuals came from out-of-bag predictions. Candidate-specific trust was defined as:

$$T(x) = \frac{1}{1 + \frac{\sqrt{u_{ep}{}^2(x) + u_{res}{}^2}}{q}} \quad (4)$$

The transformation maps predictive dispersion and residual variability to a continuous score in (0,1]. Trust decreases as either term increases and rescales the exploration contribution to acquisition. The residual scale captures variation unexplained by the predictive pipeline, including model error, omitted information and experimental variability. Here, $q > 0$ is the reference scale, expressed in the same units as both uncertainty terms. Setting q to one unit of the response scale used in the calculation recovers the original numerical transformation; this notation does not introduce target standardization. Both uncertainty terms enter the transformation on the same response scale. Response units and any target transformation therefore affect trust and the balance between predicted response and exploration. Standardizing input features does not remove this dependence on the response scale. Trust is a task- and scale-specific decision score, rather than a calibrated probability of correctness or measurement reliability.

For a Gaussian composition likelihood with draw-specific mean $\mu_{ir}$ and variance $v_r > 0$, the unnormalized inverse-likelihood importance ratio satisfies

$$log r_{ir} = \frac{1}{2} log v_r + \frac{(y_i - \mu_{ir})^2}{2v_r} + C_i \quad (5)$$

Here, r indexes posterior draws and $C_i$ is constant across draws. This identity concerns the composition layer conditional on the evidence matrix, not leave-one-out refitting of the complete pipeline. The residual MAD measures dispersion around the median residual and does not measure a common prediction bias.

Feature-strategy adaptation was separate from posterior model composition. The eight candidate strategies used mutual-information, elastic-net, neural-multiview, clustering-based, permutation-importance, attention-style, SHAP and meta-ensemble feature weights (Supplementary Table 5). For each dataset, we scored every strategy by five-fold cross-fitted $R^2$, averaged across RidgeCV and random-forest proxy models. Preprocessing and feature weights were refitted within every training fold. Figure 2d reports predictive rewards on the $R^2$ scale. For policy selection, rewards were min–max normalized within each dataset. An upper-confidence-bound policy ($\delta = 0.1$) selected one strategy for the complete evidence batch. Within each evidence fold, we refitted the preprocessing and feature-weight parameters of that selected strategy. This policy was used only for Fig. 2c,d and was distinct from trust-modulated experiment acquisition.

Figure 2e compares full SPARC with three component-removal variants and two simplified alternatives: the best single model and a non-nested k-fold evaluation. The component variants remove sparse selection, non-negativity or stability screening. The single-model alternative retains the best-performing candidate model selected using training data only. The non-nested k-fold variant examines the effect of changing the evidence-evaluation procedure. Changes in $R^2$ are reported relative to full SPARC as $100 \times (R^2_{variant} - R^2_{full})/R^2_{full}$. Definitions of the feature strategies, feature-strategy reward normalization, stability screening and variant implementations are provided in Supplementary Notes 3, 5 and 6.

**Pretrained representations and LLM comparisons**

Pretrained representations were compared on the alloy-oxide OER-stability and polymer HER tasks. For OER stability, we separately augmented the baseline descriptors with ROOST, CrabNet or Mat2Vec representations. For polymer HER, we separately augmented them with ChemBERTa, Mat2Vec or a pretrained graph convolutional network (GCN) representation.[34] Each augmented representation and its corresponding descriptor baseline were evaluated under the same leakage-controlled protocol.

GPT-4o, GPT-5.2 and GPT-5.5 received the same initial feature representations as SPARC. In the standalone condition, each LLM selected and used features and made subsequent adaptive decisions without SPARC's structured decision modules. In the hybrid condition, SPARC retained its feature-strategy and model-hypothesis libraries, while the LLM selected among the available options. Language-model configurations, prompts and logged results are reported in Supplementary Note 7.

**Learning-curve bottleneck diagnostic under a fixed representation**

Using fixed descriptors, we constructed repeated held-out learning curves for alloy-oxide OER stability and polymer HER. We used a fixed ensemble of Extra Trees, random forest and gradient boosting, distinct from the full posterior model in Fig. 3a,b. For each dataset, we generated 20 random 80:20 train–test splits. Within each split, we evaluated progressively larger nested training subsets against the same test set. We refitted preprocessing for every training subset and averaged predictions across the three regressors.

We fitted the monotonic envelope of the mean held-out $R^2$ values with the saturating response:

$$R^2(N) = R^2_{ceiling} - Aexp\left[-\frac{N - N_{min}}{\tau}\right] \quad (6)$$

Here, $N$ is the training-set size, $N_{min}$ is the smallest evaluated size, $R^2_{ceiling}$ is the fitted finite-range ceiling and $\tau$ is the fitted saturation scale. We defined the relative bottleneck index as:

$$B(N) = 1 - exp\left[-\frac{N - N_{min}}{\tau}\right] \quad (7)$$

For an increasing saturating curve, $A > 0$ and $\tau > 0$; A is the fitted improvement from Nmin to the asymptote. The relative index can equivalently be written as

$$B(N) = \frac{R^2(N) - R^2(N_{min})}{{R_{ceiling}}^2 - R^2(N_{min})} \quad (8)$$

B(N) measures the fraction of fitted improvement achieved between the smallest training size and the asymptote. At B(N) = 0.5, $N^* = N_{min} + \tau ln2$, and the fitted marginal gain is half that at $N_{min}$. We interpreted B(N) as a relative measure of saturation over the sampled size range. It does not estimate a universal sample-size threshold, an irreducible performance limit or a bound on SPARC. We estimated 90% bootstrap intervals from 2,000 resamples of the repeated learning curves. Full split-level curves and fitted parameters are provided in Supplementary Fig. 4 and Supplementary Table 7.

When A = 0, the improvement ratio is undefined, and a half-improvement transition is not identifiable.

**Finite-pool candidate optimization**

For alloy oxides, candidate spaces used precursor identities and composition patterns from the dataset. For polymers, candidate spaces used decomposition and compatible recombination of scaffold, substituent and polymerizable-site fragments (Fig. 3e,g). The analyses in Fig. 3f,h were retrospective comparisons within fixed finite pools of archived candidate outcomes.

Iterative acquisition used a heterogeneous bootstrap ensemble to update predictive means and dispersion as candidate outcomes were revealed. The learner library contained two Ridge regressors, RidgeCV, two random-forest regressors and one gradient-boosting regressor. We cycled these six learners across 150 bootstrap resamples drawn with replacement and refitted preprocessing within every resample. The mean and standard deviation across replicate predictions supplied the candidate prediction and predictive dispersion, respectively. We estimated the residual scale as the normal-consistent median absolute deviation of out-of-bag residuals. Candidate-specific trust combined this residual scale with the bootstrap predictive standard deviation.

Candidates were ranked using a trust-modulated confidence bound:

$$a_T(x) = s\mu(x) + \beta u_{boot}(x)T(x) \quad (9)$$

Here, $\mu(x)$ and $u_{boot}(x)$ denote the bootstrap mean and standard deviation, respectively; $\beta = 1.96$, and s sets the objective direction. We maximized aT(x), using $s = +1$ for HER. For Ru-dissolution minimization, this was equivalent to minimizing $\mu(x) - \beta u_{boot}(x)T(x)$ . Trust continuously moderated the exploration bonus while retaining every unmeasured candidate in the ranking. For a fixed residual scale, the moderated bonus increased with predictive dispersion but approached a finite limit.

The coefficient β is an exploration parameter rather than a coverage guarantee. With the reference scale q defined above, the exploration bonus tends to βq as the predictive standard deviation increases without bound at fixed residual scale.

All plotted strategies were initialized from the same 20 labelled experiments for Ru dissolution and the same 10 labelled samples for polymer HER. Evaluation counts included initialization. Discovery was recorded at the first evaluation whose best-so-far value lay within 0.1% of the final pool optimum. Learner configurations and acquisition parameters are provided in Supplementary Note 9.

**Controlled extrapolation**

We evaluated controlled extrapolation on the formate-oxidation and OER-activity datasets. Splits were defined from the input variables only, without responses. For formate oxidation, the first principal component of the standardized inputs was computed over the complete fixed candidate pool, and candidates above its 70th percentile formed the extrapolation pool; an 80th-percentile split was used as a sensitivity check (Supplementary Note 10). For OER activity, candidates at cobalt levels 1–3 formed the initial region and those at levels 4–7 the extrapolation pool. This produced 214 initial-region and 92 extrapolation candidates for formate oxidation, and 406 initial-region and 56 extrapolation candidates for OER activity. The construction tested expansion beyond the initial ranges of the same experimental variables rather than transfer to an unrelated chemistry. Apart from the formate-oxidation split, principal-component projections were used only for visualization; model fitting used the original input variables.

For controlled extrapolation, SPARC combined three Gaussian-process experts: an isotropic Matérn-5/2 kernel, an automatic-relevance-determination radial-basis-function kernel and a sparse-prior Gaussian process. Expert weights were derived from three-fold cross-validated predictive log likelihoods. The mixture predictive variance included both within-expert variance and between-expert disagreement. We calculated trust from the mixture predictive standard deviation and a leakage-controlled residual scale.

Writing $\pi_m$ for the non-negative expert weights with $\Sigma_m \pi_m = 1$, and $\mu_m(x)$ and $v_m(x)$ for each expert's predictive mean and variance, the mixture moments are

$$\mu(x) = \sum_m \pi_m \mu_m(x), u_{mix}{}^2(x) = \sum_m \pi_m [v_m(x) + (\mu_m(x) - \mu(x))^2] \quad (10)$$

These are moments of the mixture distribution. Combining its standard deviation with a residual scale defines a decision score, not a calibrated decomposition of total prediction error.

We ran four acquisition rounds for each extrapolation pool. Each round selected 23 candidates for formate oxidation and 14 candidates for OER activity. Batches were selected by trust-modulated Thompson sampling:

$$a_T(x) = s[\mu(x) + u_{mix}(x) T(x) z] \quad (11)$$

Here, $\mu(x)$ and $u_{mix}(x)$ denote the three-expert mixture mean and standard deviation, $z$ was drawn from a standard normal distribution, and $s$ sets the objective direction. Trust scaled the stochastic exploration term. At each round, we selected the entire batch before revealing any candidate responses. We then added these measurements to the training set and refitted all experts and mixture weights before the next round.

The score is maximized, with s = +1 for response maximization and s = −1 for response minimization. In the reported analyses, the standard normal perturbation was drawn independently for each candidate (Supplementary Note 4).

We defined the uncertainty response as the ratio of mean predictive uncertainty in the extrapolation pool to that in the initial region. Ranking was evaluated from acquisition order at complete-batch cutoffs. For cutoff fraction K, we considered the earliest acquired K fraction of the extrapolation pool. Precision@K was the fraction of these candidates belonging to the pool's true top-K fraction. The complete-batch cutoffs were 25%, 50%, 75% and 100% of each pool. Step displays summarize the evaluated cutoffs; they do not resolve ordering within a batch. Candidate outcomes were revealed only after each complete batch was selected. The expected precision of random selection at cutoff K is K. Per-round hit rate was the fraction of each selected batch belonging to the true top quarter of the original extrapolation pool. Error bars show ±1.96 binomial standard errors, $\sqrt{[p(1 - p)/n]}$, for hit fraction p and batch size n. These are nominal intervals for batch hit fractions, not variability across repeated acquisition runs.

Equivalently, if $A_k$ is the acquired set and $H_k$ is the true top-K set, each of size KN, then

$$Precision@K = \frac{|A_K \cap H_K|}{KN}, E_{random}[Precision@K] = K \quad (12)$$

**Automated OER validation and measurement verification**

The prospective study used an automated electrodeposition and electrochemical characterization workflow adapted from our previously reported platform.[12,35] Synthesis and electrochemical testing followed our previously reported procedures;[12,35] campaign-specific conditions are listed in Supplementary Note 11. The initial dataset comprised 125 independent alkaline OER experiments spanning seven transition-metal cations, two anion chemistries and five synthesis variables: deposition temperature, current, time, complexing-agent volume and precursor pH. The response was the overpotential required to reach 10 mA $cm^{-2}$.

We fitted SPARC to the 125 initial experiments and selected 11 conditions for the next round. Recommendations used the heterogeneous bootstrap ensemble and the trust-modulated lower confidence bound for overpotential minimization. We constructed the batch greedily after standardizing all features. A candidate was excluded from the current batch if its standardized feature-space distance from an already selected condition was below $0.15\sqrt{d}$, where d is the number of standardized features. All 11 recommendations were fixed before their experimental outcomes were measured.

We assessed verification priority using four diagnostics and reported trust separately. Conformal nonconformity was flagged at an empirical P value below 0.10.[37] The Ridge-proxy influence score was the absolute difference between a full-fit Ridge prediction and the leave-one-out prediction of the model ensemble for the same record; observations in the upper 10% were flagged. Isolation Forest used a contamination fraction of 0.10.[38] Mahalanobis distance was calculated under Ledoit–Wolf shrinkage covariance and flagged at a $\chi^2$ P value below 0.05.[39] Measurements with at least two diagnostic flags entered the priority-remeasurement group; trust was not an additional binary flag. Controls had no diagnostic flags and were matched to the priority group in feature space. All group assignments were recorded before repetition.

We repeated eight unflagged control measurements and seven priority measurements. Repeatability was quantified as the absolute difference between the original and repeated overpotentials. We report individual deviations and descriptive group means and medians. The verification endpoint was the observed distribution of original–repeat differences in the prespecified groups. Repeated values were evaluated independently of acquisition; training-time soft gating and repeat-fed acquisition were not used. Diagnostic settings, matching procedures and experimental procedures are provided in Supplementary Notes 11 and 12.

The Supplementary Information reports the complete model definitions, hyperparameters, prior distributions and posterior-sampling settings (Supplementary Notes 2–4), feature-strategy implementations (Supplementary Note 5), LLM configurations (Supplementary Note 7), acquisition parameters (Supplementary Notes 4 and 9), experimental procedures (Supplementary Note 11), and software versions, settings and random seeds (Supplementary Note 13 and Supplementary Table 3).

**Data availability**

The eight benchmark datasets are available from their original publications, which are listed in Supplementary Table 1. Processed datasets and figure source data are available from the SPARC repository at https://github.com/SissiFeng/SPARC. The data underlying the Supplementary Figures and Tables are provided as Supplementary Data 1–4. Source data are provided with this paper.

**Code availability**

The code used in this work is available from the SPARC repository at https://github.com/SissiFeng/SPARC under the Apache-2.0 licence. The version used in this study (v2.5.3rc4) is archived at Zenodo (https://doi.org/10.5281/zenodo.24201692). The repository includes the implementations used for the trust-modulated confidence-bound and Thompson-sampling decision analyses, as well as the learning-curve bottleneck analysis. Reproducibility information, software dependencies and environment specifications are provided in the repository.

**Acknowledgements**

This research was undertaken thanks in part to funding provided to the University of Toronto's Acceleration Consortium from the Canada First Research Excellence Fund (grant number CFREF-2022-00042).

## Author contributions

S.F. developed the SPARC framework and its software implementation and performed the computational analyses. Y.H. and R.Z. performed the automated experiments, and Y.H. also reviewed the code. Y.L. contributed to the code development. J.B. contributed to discussions on model development. A.S.Z. contributed to discussions on the electrochemical platform. B.S., J.H.-S., V.B. and A.A.-G. revised the manuscript. Y.B. conceived the project and supervised the work. All authors reviewed and approved the final manuscript.

## Competing interests

The authors declare no competing interests.

# Supplementary Information

## SPARC: Evidence-guided adaptive model composition for materials discovery

Sissi Feng[1,†], Yuchen Han[1,2,†], Runze Zhang[1,2,†], Jiaru Bai[3,4], Yixi Li[1], Brandon Sutherland[1], Ali Shayesteh Zeraati[1], Jason Hattrick-Simpers[1,2], Varinia Bernales[1,3,5], Alán Aspuru-Guzik[1,2,3,4,5,6,7,8,9], Yang Bai[1,*]

[1]*Acceleration Consortium, University of Toronto, Toronto, Ontario, Canada.*

[2]*Department of Materials Science and Engineering, University of Toronto, Toronto, Ontario, Canada.*

[3]*Department of Chemistry, University of Toronto, Toronto, Ontario, Canada.*

[4]*Vector Institute for Artificial Intelligence, Schwartz Reisman Innovation Campus, Toronto, Ontario, Canada.*

[5]*Department of Computer Science, University of Toronto, Toronto, Ontario, Canada.*

[6]*Department of Chemical Engineering & Applied Chemistry, University of Toronto, Toronto, Ontario, Canada.*

[7]*Institute of Medical Science, University of Toronto, Toronto, Ontario, Canada.*

[8]*Canadian Institute for Advanced Research (CIFAR), Toronto, Ontario, Canada.*

[9]*NVIDIA, Toronto, Ontario, Canada.*

[†]*These authors contributed equally.*

[*]*Correspondence: yangb.bai@utoronto.ca*

This Supplementary Information documents the algorithmic definitions, implementation settings and additional experiments behind SPARC (Sparse Predictive Aggregation of Regression Components). It is organized in the order of the main-text argument. Part I specifies how evidence, composition, uncertainty and trust are computed. Part II examines predictive behaviour: benchmark ablations, representation and language-model comparisons and the limits imposed by the available data. Part III follows the same models into decisions: finite-pool search and controlled extrapolation. Part IV reports the automated oxygen-evolution-reaction (OER) campaign, its prospective prediction record and measurement verification. Part V describes software and data.

The document contains 13 Supplementary Notes, 8 Supplementary Figures, 8 Supplementary Tables and 4 Supplementary Data files. Counts reported for different analyses refer to different analysis populations (Supplementary Note 1).

**Contents**

## Part I. Framework and implementation

### Supplementary Note 1. Study design, tasks and evaluation units

**Benchmarks.** The predictive benchmarks cover formate oxidation, organic laser emitters, inorganic synthesis, OER activity, photovoltaics, polymer hydrogen evolution reaction (HER), OER stability and battery performance.[1–7] The underlying studies span robotic electrocatalyst discovery, molecular design and high-throughput electrochemistry (Supplementary Table 1). Molecular tasks use fingerprints and physicochemical descriptors; inorganic tasks use composition and precursor information; process variables describe synthesis or testing conditions where they were recorded. Each task retains the descriptors native to its domain, so representation differences between tasks are part of the benchmark rather than a confounder to be removed.

**Evaluation unit.** These tasks have different responses, units and experimental designs. Literature-reported scores are therefore quoted only as context for each task, under the protocol of the source study, and not as matched-split baselines. Comparisons with reimplemented methods use identical outer test observations. In the predictive benchmarks, outer leave-one-out evaluation refits every data-dependent operation on the outer training partition: imputation, categorical encoding, scaling, zero-variance filtering, any enabled feature-strategy selection, hyperparameter selection, stability screening, evidence construction and posterior composition. A held-out observation therefore evaluates the complete fitted pipeline. For regression, the coefficient of determination is computed once from the assembled outer predictions,

$$R^2 = 1 - \frac{\sum_{i=1}^{n}(y_i - \hat{y}_i^{\text{out}})^2}{\sum_{i=1}^{n}(y_i - \bar{y})^2}, \qquad \text{(S1)}$$

rather than by averaging single-observation fold scores, which are undefined. Mean absolute error (MAE) and root-mean-square error (RMSE) retain the response units, and Spearman's $\rho$ summarizes rank agreement. Negative $R^2$ values are reported as obtained, because they identify predictions worse than the constant-mean reference.

**Analysis populations.** The supplementary analyses use their own evaluation units, listed in Supplementary Table 2. The fixed-pipeline learning curves use 99 polymer records and 160 OER-stability records. Controlled extrapolation uses fixed candidate pools of 306 formate-oxidation and 462 OER-activity records. The alkaline OER campaign starts from 125 experiments and evaluates 11 further recommendations. These populations overlap in scientific domain but serve different analyses.

**Reported analyses and archived software.** Supplementary Notes 2–12 describe the settings used to obtain the results reported here and in the main text. The archived software (v2.5.3rc4) implements these settings, with the two exceptions listed in Supplementary Table 3.

**Supplementary Table 1 | Published sources associated with the benchmark tasks.** Task names follow the main text. Source protocols are not treated as matched-split comparisons.

| Task | Material or experimental scope | Source study |
|---|---|---|
| Formate oxidation | Multi-element electrocatalysts | Zhang et al.[1] |
| Organic laser emitters | Organic molecular laser emitters | Strieth-Kalthoff et al.[2] |
| Inorganic synthesis | Sol–gel synthesis of mixed-metal oxides (gel formation) | Bai et al.[3] |
| OER activity | High-entropy catalysts | Zhou et al.[4] |
| Photovoltaics | Hole-transport materials for perovskite cells | Wu et al.[5] |
| Polymer HER | Organic polymer photocatalysts | Bai et al.[6] |
| OER stability | Alloy-oxide catalyst stability | Bai et al.[3] |
| Battery performance | Electrochemical screening for battery materials | Lin et al.[7] |

**Supplementary Table 2 | Analysis populations and implementations.** Rows describe distinct analyses; their counts are not additive. ID, in-distribution; OOD, out-of-distribution; GP, Gaussian process; PC1, first principal component.

| Analysis | Evaluation unit | Implementation and scope | Note |
|---|---|---|---|
| Predictive benchmarks | Outer held-out record | 12-model Bayesian composition | 2, 3, 6 |
| Representation and LLM comparison | Outer held-out record | Main-text Fig. 3a,b point estimates | 7 |
| Logged GPT-5.5 standalone runs | 99 or 160 held-out rows | Standalone five-shot interface, five folds | 7 |
| Fixed-pipeline learning curves | Repeat × training size | Three-model tree ensemble, 20 splits | 8 |
| Controlled extrapolation | Fixed ID/OOD split | Three-GP mixture; cobalt-level and PC1 splits | 10 |
| Automated OER campaign | Experiment | 125 initial, 11 recommended | 11 |

**Supplementary Table 3 | Differences between the reported analyses and the archived software.** All other settings described in this document are those of the archived software (v2.5.3rc4).

| Component | Reported analyses | Archived software |
|---|---|---|
| Third Gaussian-process expert | Sparse-prior GP | Radial-basis-function kernel with length scales set from random-forest importance |
| Influence diagnostic | Full-fit Ridge prediction minus the leave-one-out prediction of the model ensemble | Full-fit Ridge prediction minus the prediction of the same Ridge model refitted without the record |

## Supplementary Note 2. Model library and hyperparameter selection

SPARC assigns composition weights to individual predictors drawn from a library of 12 models in four structural families (Supplementary Table 4). Linear models (Ridge, Lasso, elastic net, Bayesian ridge and Huber regression) express responses that change linearly with each descriptor. Tree-based nonlinear models (random forest, Extra Trees and gradient boosting) represent thresholds and interactions. Locally smooth models (radial-basis-function support-vector regression, distance-weighted k-nearest neighbours and a Matérn-5/2 Gaussian process) assume that similar inputs give similar responses. A monotonic histogram gradient-boosting regressor expresses responses that move consistently in one direction with selected descriptors. The family labels identify modelling assumptions; they do not imply that each predictor represents a separate physical mechanism, and weights are assigned to individual models rather than to families.

**Hyperparameter selection.** All analyses used the fixed hyperparameters listed in Supplementary Table 4, held constant across tasks and outer folds. Bayesian ridge regression estimates its own regularization, and the Gaussian-process kernel hyperparameters were optimized by marginal likelihood with two random restarts.

**Monotonic constraints.** The monotonic regressor imposes an increasing or decreasing constraint on each feature whose Spearman correlation with the response, computed in the training state, is at least 0.3 in magnitude; other features are unconstrained. Constraint directions are therefore re-estimated in every outer fold and do not use the held-out response.

**Supplementary Table 4 | Twelve-model library.** The same fixed settings apply in every fit. Unspecified arguments follow the installed library defaults (scikit-learn 1.6.1). Random seed 42 is used where applicable.

| Family | Model | Settings |
|---|---|---|
| Linear | Ridge | $\alpha = 1$ |
| | Lasso | $\alpha = 10^{-3}$; 5,000 iterations |
| | Elastic net | $\alpha = 10^{-3}$; $\ell_1$ ratio 0.5; 5,000 iterations |
| | Bayesian ridge | Library defaults |
| | Huber | $\varepsilon = 1.35$; 500 iterations |
| Tree-based | Random forest | 200 trees; maximum depth 10 |
| | Extra Trees | 200 trees; maximum depth 10 |
| | Gradient boosting | 150 stages; maximum depth 4 |
| Locally smooth | RBF support-vector regression | $C = 10$; $\varepsilon = 0.1$ |
| | k-nearest neighbours | Five neighbours (clipped to training size); distance weights |
| | Gaussian process | Constant × Matérn-5/2; initial length scale 1; diagonal term $10^{-2}$; normalized response; two optimizer restarts |
| Monotonic | Monotonic gradient boosting | Histogram boosting; 200 iterations; depth 3; learning rate 0.05; minimum leaf 5; directions from Spearman $\|\rho\| \geq 0.3$ |

## Supplementary Note 3. Evidence matrix, stability screening and posterior composition

**Evidence matrix.** For a training set of n labelled observations and M candidate models, the evidence matrix is

$$P_{im} = f_m^{(-i)}(x_i), \qquad P \in \mathbb{R}^{n\times M}, \qquad \text{(S2)}$$

where $f_m^{(-i)}$ is fitted without observation i. Every entry is thus a genuine out-of-sample prediction. Within an outer evaluation fold, P is built from the outer training records only. After composition has been estimated, the retained predictors are refitted on that training set before predicting the outer test record or an unmeasured candidate. The evidence used to decide model influence and the outer predictions used to judge generalization therefore play separate roles.

**Stability screening.** Support-set screening follows the principle of repeated selection under data perturbation.[10] Fifty subsamples, each containing 80% of the training records, are drawn without replacement. Within each subsample, leave-one-out evidence is constructed and a non-negative Lasso ($\alpha = 0.01$, with intercept) is fitted to the evidence and response after both are divided by the normal-consistent median absolute deviation (MAD) of the subsample responses. A model is selected in a subsample if its coefficient exceeds $10^{-5}$. Models selected in at least 60% of subsamples enter the active set $\mathcal{A}$. Screening is omitted for training sets with fewer than 20 records, and if no model reaches the threshold the five most frequently selected models are retained. Screening is the only step that sets weights exactly to zero.

**Posterior composition.** For $m \in \mathcal{A}$, the composition is

$$a_m \mid s_m \sim \mathcal{N}(0, s_m^2), \quad s_m \mid \rho, c_m \sim \text{Half-Cauchy}(0, \rho c_m), \quad \rho \sim \text{Half-Cauchy}(0,1), \qquad \text{(S3)}$$

$$w_m = \frac{|a_m|}{\sum_{j \in \mathcal{A}} |a_j|}, \qquad y \mid P, w, \sigma^2 \sim \mathcal{N}(Pw, \sigma^2 I), \qquad \sigma^2 \sim \text{Inverse-Gamma}(1,1). \qquad \text{(S4)}$$

The weights are non-negative and sum to one. The positive factor $c_m$ is shared by models of the same family and defaults to one. Hierarchical half-Cauchy scales are also used in sparse axis-aligned subspace Bayesian optimization,[11] although here they act on coefficients whose magnitudes are normalized into composition weights. A common multiplicative scale cancels under normalization, so a family-specific prior preference requires unequal relative family scales; with uniform $c_m$, removing family priors leaves the model unchanged. The posterior is sampled with the No-U-Turn sampler[12] using a non-centred parameterization, four chains, 1,000 warm-up iterations, 2,000 retained draws per chain and target acceptance probability 0.85. Sampling was accepted when every normalized weight had $\hat{R} < 1.01$ and an effective sample size above 400. If either criterion failed, sampling was repeated once with 2,000 warm-up iterations. Posterior-mean weights give the point prediction.

**Composition uncertainty.** All candidates share the posterior over weights at a given training state. For posterior draw r,

$$\mu_r(x) = \sum_{m \in \mathcal{A}} w_{mr} f_m(x), \qquad u_{\text{ep}}^2(x) = f(x)^{\top} \operatorname{Cov}(w \mid P, y) f(x), \qquad \text{(S5)}$$

and $u_{\text{ep}}(x)$ is computed as the standard deviation of $\mu_r(x)$ across draws. This is uncertainty in the composed mean conditional on the fitted response surfaces. It differs from disagreement among model predictions at a fixed weight vector and does not include every source of within-model parameter uncertainty. Because $f(x)$ varies across candidates while the weight posterior does not, the same posterior produces small uncertainty where retained models agree and large uncertainty where they diverge.

**Residual scale.** Variation not explained by the composition is summarized by

$$u_{\mathrm{res}} = 1.4826\,\mathrm{median}_i\left|e_i - \mathrm{median}_j\left(e_j\right)\right|, \qquad \text{(S6)}$$

computed from residuals of a leave-one-out approximation at the composition layer. If draw r has mean $\mu_{ir}$ and variance $v_r$, its unnormalized inverse-likelihood ratio is

$$\log r_{ir} = 1/2\,\log v_r + \frac{(y_i - \mu_{ir})^2}{2v_r} + C_i, \qquad \text{(S7)}$$

with $C_i$ constant across draws. The normalized ratios are capped at 10% and renormalized, a form of truncated importance sampling.[8] This calculation is conditional on P and does not replace outer refitting of the full pipeline. The MAD measures dispersion around the median residual; a common prediction bias is not captured by it.

## Supplementary Note 4. Uncertainty, trust and acquisition rules

**Trust.** For a candidate-dependent dispersion $u(x)$ and residual scale $u_{\mathrm{res}}$, trust is

$$T(x) = \left[1 + \frac{\sqrt{u(x)^2 + u_{\mathrm{res}}^2}}{q}\right]^{-1}, \qquad q > 0. \qquad \text{(S8)}$$

The dispersion is $u_{\mathrm{ep}}$ for the Bayesian composition, the bootstrap standard deviation $u_{\mathrm{boot}}$ for finite-pool and laboratory acquisition, and the mixture standard deviation $u_{\mathrm{mix}}$ for controlled extrapolation. The reference scale q has the units of the response. All analyses set q to one response unit (for example, 1 mV for overpotential). Trust lies in (0, 1], decreases as either uncertainty term grows, and is a task- and scale-specific decision score, not a calibrated probability that a prediction or measurement is correct.

**Confidence-bound acquisition.** Following upper-confidence-bound selection,[13] candidates are ranked by maximizing

$$a_T(x) = s\,\mu(x) + \beta\,u(x)\,T(x), \qquad \beta = 1.96, \qquad \text{(S9)}$$

with s = +1 for maximization and s = −1 for minimization. At fixed $u_{\mathrm{res}}$, the exploration bonus increases with $u(x)$ but is bounded above by βq. The choice of q therefore controls how much exploration is possible. Trust moderates the bonus continuously and does not remove any candidate from consideration.

**Batch Thompson sampling.** For controlled extrapolation, each batch is ranked by

$$a_T(x) = s\left[\mu(x) + T(x)\left(\tilde{f}(x) - \mu(x)\right)\right], \qquad \text{(S10)}$$

where $\tilde{f}$ is a randomized draw of the response, obtained by adding to the mixture mean an independent standard-normal perturbation $z_x$ for each candidate, scaled by $u_{\mathrm{mix}}(x)$.[14]

**Supplementary Note 5. Feature-strategy adaptation**

Feature weighting and posterior model weighting act at different stages. The eight strategies in Supplementary Table 5 transform the input representation before the evidence matrix is assembled; posterior composition then weights models on that representation. Within each outer training set, an upper-confidence-bound policy ($\delta = 0.1$) selects among the strategies. Each strategy is scored by five-fold cross-fitted $R^2$, averaged over RidgeCV ($\alpha \in \{0.1, 1, 10\}$) and a 60-tree random forest of depth 6, with preprocessing and feature weights refitted in every fold. Rewards are min–max normalized within the training set, and the strategy with the highest normalized reward is selected for the complete evidence batch, after which its weights are refitted inside each evidence fold. A strategy whose scoring fails is recorded as missing, which is the meaning of a grey cell in main-text Fig. 2d; it is not a zero score. Scoring is deterministic for a given training set, so the policy reduces to selecting the highest-scoring strategy.

The names 'neural-multiview' and 'attention-style' refer to weighting heuristics. The former sums the magnitudes of a fitted perceptron's input-layer weights; the latter further scales these by the feature–response association. Neither establishes learned chemical views or mechanistic attention, and a feature weight or attribution is not evidence of a causal effect on the response.

**Supplementary Table 5 | Feature-weighting strategies.** All supervised quantities are fitted on the corresponding training partition. Weights are normalized to mean one before they rescale standardized features.

| Strategy | Construction |
|---|---|
| Mutual information | Univariate mutual-information regression scores |
| Elastic net | Absolute coefficients after standardization; $\alpha = 1$, $\ell_1$ ratio 0.5, 2,000 iterations |
| Neural-multiview | Sum of absolute input-layer weights of a one-hidden-layer perceptron; width min(64, max(8, $\lfloor d/2 \rfloor$)); 300 iterations with early stopping |
| Clustering | Inverse mean within-cluster feature variance after standardization; k-means with min(max(2, $\lfloor n/5 \rfloor$), 10) clusters and five initializations |
| Permutation importance | Non-negative mean decrease in training $R^2$ over five permutations; random forest with 50 trees and depth 5 |
| Attention-style | Neural input-weight magnitude multiplied by one plus the absolute standardized feature–response association |
| SHAP | Mean absolute TreeSHAP attribution[15] for a random forest with 100 trees and depth 6 |
| Meta-ensemble | Mean of separately min–max normalized mutual-information, permutation and SHAP scores |

# Part II. Predictive evidence

## Supplementary Note 6. Component ablation on the published benchmarks

Main-text Fig. 2e compares full SPARC with three component-removal variants and two simplified alternatives. For a positive full-model $R^2$, the reported quantity is

$$\Delta R^2_{\mathrm{rel}} = 100\,\frac{R^2_{\mathrm{variant}} - R^2_{\mathrm{full}}}{R^2_{\mathrm{full}}}. \qquad \text{(S11)}$$

Removing sparse selection lowered $R^2$ on all eight tasks, with relative losses ranging from 0.2% on battery performance to 20.0% on formate oxidation. Removing non-negativity also lowered $R^2$ on all eight tasks, with losses ranging from 0.6% on polymer HER to 20.0% on formate oxidation. Removing stability screening lowered $R^2$ on seven tasks, with losses of 6.0–33.0%, while inorganic synthesis showed no change at the reported precision. The largest loss from removing stability screening occurred on photovoltaics. The non-nested k-fold variant yielded relative losses of 4.1–32.7% across the eight tasks. Selecting only the best single model lowered $R^2$ on every task, with losses ranging from 5.0% on OER stability to 40.5% on polymer HER; the losses on organic laser emitters and battery performance were 23.4% and 18.0%, respectively. These point estimates support a predictive benefit from composition over the selected single-model alternative across the benchmark suite, with task-dependent contributions from the individual components.

The definitions of the variants are as follows. *No sparse selection* replaces the hierarchical half-Cauchy local scales with fixed scales. *No non-negativity* allows signed coefficients normalized by their signed sum, with exact integration over sign patterns. *No stability screening* composes all twelve models. *Best single model* uses the model with the lowest training-only leave-one-out error. The *non-nested k-fold* variant tunes, screens and composes once on all observations using five-fold evidence and applies these weights to the same outer-held-out base-model predictions used by the nested variants, so it is scored on the same observations. The values in main-text Fig. 2e are single-run point estimates; no between-run uncertainty is inferred from them.

Supplementary Fig. 1 compares the outer held-out predictions of full SPARC with the measured responses for the eight benchmark tasks; these predictions give the values in main-text Fig. 2b. Inorganic synthesis is a binary classification task (whether a sol–gel reaction forms a gel), for which the reported value is classification accuracy rather than $R^2$.

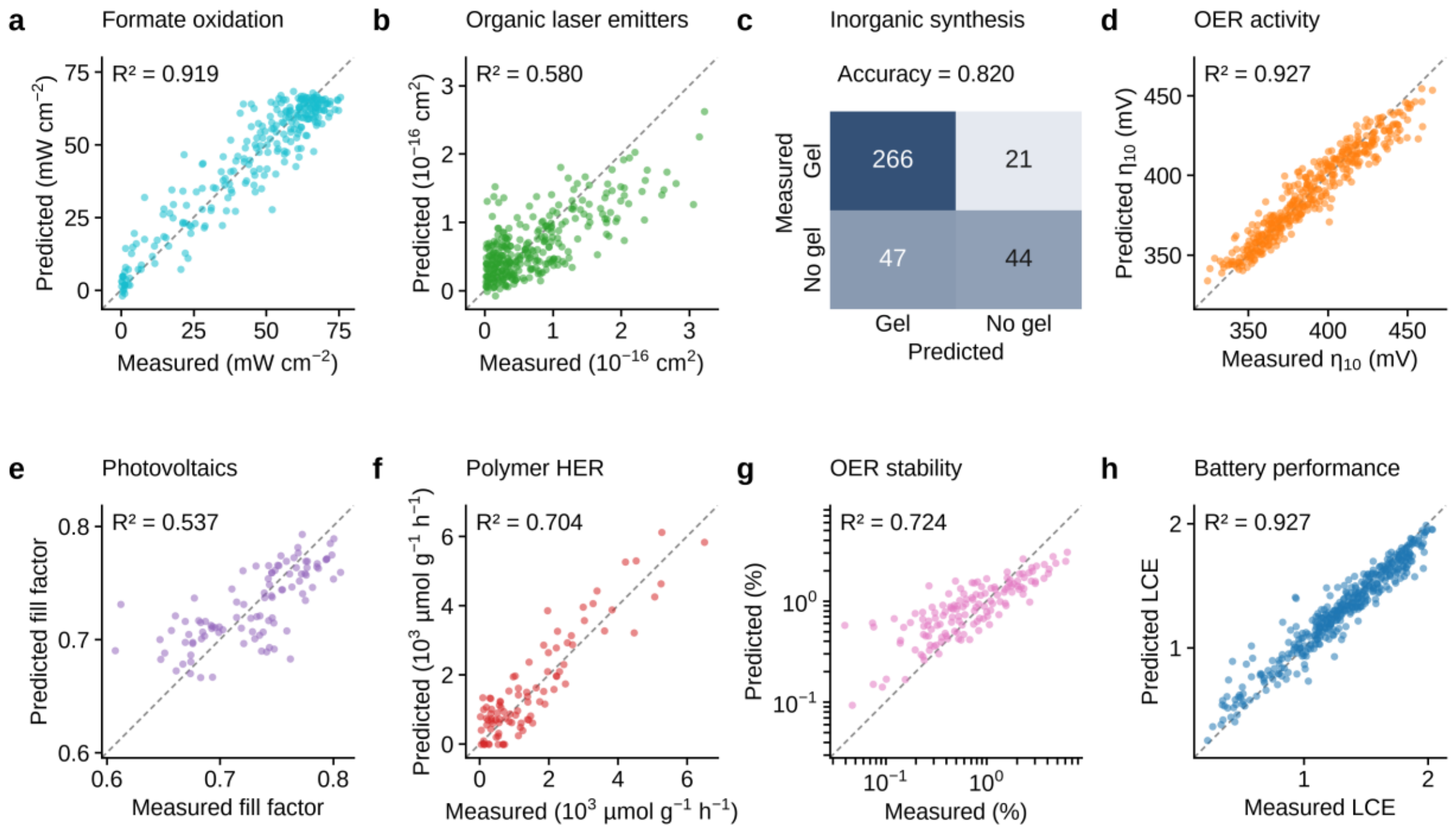

**Supplementary Fig. 1 | Outer held-out predictions of full SPARC on the eight benchmark tasks. a**, Formate oxidation (power density). **b**, Organic laser emitters (emission gain cross-section). **c**, Inorganic synthesis, shown as a confusion matrix of measured and predicted sol–gel outcome (gel formed or not); shading gives the fraction of each measured class. **d**, OER activity. **e**, Photovoltaics. **f**, Polymer HER. **g**, OER stability, on logarithmic axes. **h**, Battery performance. In a, b and d–h, each point is one record and the dashed line marks parity. Each panel reports the value shown in main-text Fig. 2b: $R^2$ for the regression tasks and classification accuracy for inorganic synthesis. Colours follow main-text Fig. 2a. $\eta_{10}$, overpotential at 10 mA $cm^{-2}$; LCE, logarithmic Coulombic efficiency, $-\log_{10}(1 - CE)$.

**Supplementary Note 7. Pretrained representations and language models**

**Pretrained embeddings.** Main-text Fig. 3a,b asks whether representations learned on large external corpora improve small experimental datasets. The OER-stability descriptors were augmented separately with ROOST, CrabNet or Mat2Vec embeddings, and the polymer HER descriptors with ChemBERTa, Mat2Vec or a pretrained molecular graph convolutional network (GCN) representation. Pretrained encoders were applied in evaluation mode without parameter updates: ChemBERTa embeddings were frozen, Mat2Vec representations were element-composition averages of the published embeddings, and the GCN encoder was loaded from its pretrained checkpoint. Each representation was concatenated with the baseline descriptors and passed through the same nested pipeline, outer partitions and random seeds as its descriptor baseline (main-text Fig. 3a,b), with all target-dependent preprocessing, feature selection and model fitting restricted to the training partitions. All six augmentations lowered $R^2$: by 10.9–17.5% for OER stability and by 27.7–37.1% for polymer HER. With 160 and 99 records, adding high-dimensional embeddings increases the variance of every model more than it adds useful signal. Domain descriptors of a few dozen dimensions remained the better representation at these dataset sizes.

**Language-model interfaces.** Two ways of using a large language model (LLM) were compared. In the *standalone* condition, the LLM receives the initial representation and makes the modelling decisions itself, without SPARC's structured modules. In the *hybrid* condition, the LLM selects among SPARC's available feature strategies and model hypotheses, and the composition proceeds as usual. In the main-text comparison with GPT-5.2, the hybrid condition raised OER-stability $R^2$ from 0.724 to 0.750 (+3.6%) and lowered polymer $R^2$ from 0.704 to 0.684 (−2.8%). These are single point estimates and do not identify a repeated-run distribution, so they do not establish a general advantage for an LLM selector.

**Language-model configurations.** Each request record stores the model identifier, system and user messages, decoding parameters and required output structure; decoding settings are specified per run rather than inferred from the model family. In the standalone mode, the LLM first returns a non-empty set of distinct feature indices and then one finite numerical prediction per query row. In the hybrid mode, it returns one of the supplied feature-strategy names and a non-empty set of supplied model names, which SPARC then fits and combines on the training observations. Training responses may appear as labelled examples, whereas held-out records contribute descriptors only, and malformed responses are not converted into predictions.

**GPT-4o polymer baseline.** Requests used the API model identifier gpt-4o with temperature = 0 and max_tokens = 500; top_p, seed and penalty parameters were not supplied. Five-fold evaluation used five training examples per prompt. The system message was 'You are a materials science expert. Respond only with valid JSON.' The user message specified the task (photocatalytic hydrogen evolution under TEA/MeOH/$H_2O$ conditions) and the inputs (SMILES of monomers A and B, Pd content, optical gap, photoluminescence lifetime and transmittance), listed the five training examples in the same format with their HER values, and requested a JSON object containing a numerical her_prediction in $\mu$mol g$^{-1}$ h$^{-1}$ and a brief reasoning field. JSON output was requested in the prompt rather than enforced through an API response-format argument. The full prompt template is:

You are a materials science expert specializing in photocatalytic hydrogen evolution. Given a polymer photocatalyst with: Monomer A SMILES (aromatic building block); Monomer B SMILES (aromatic building block); Pd content (%); Optical gap (eV); PL lifetime (ns); Transmittance (%). Predict the hydrogen evolution rate (HER) in μmol/(g·h) under TEA/MeOH/H2O conditions. Here are some training examples: {five formatted training-example blocks} Now predict for this NEW polymer: Monomer A: {unit_a_smiles} Monomer B: {unit_b_smiles} Pd content: {pd_content_pct:.4f}% Optical gap: {optical_gap_meas_ev:.2f} eV PL lifetime: {pl_lifetime_ns:.2f} ns Transmittance: {transmittance_tea_meoh_water_pct:.2f}% Respond ONLY with a JSON object containing: {"her_prediction": <number>, "reasoning": "<brief explanation>"} Your prediction:

Each training-example block has the form 'Example {i + 1}: Monomer A: {unit_a_smiles} Monomer B: {unit_b_smiles} Pd content: {pd_content_pct:.4f}% Optical gap:

{optical_gap_meas_ev:.2f} eV PL lifetime: {pl_lifetime_ns:.2f} ns Transmittance: {transmittance_tea_meoh_water_pct:.2f}% → HER: {her_value:.2f} µmol/(g·h)'. Braces denote substituted values.

**Logged GPT-5.5 experiments.** Later runs saved complete per-record predictions for the standalone interface on polymer HER and OER stability (Supplementary Figs 2 and 3; Supplementary Table 6). These runs used GPT-5.5 and therefore form a separate experiment from the GPT-5.2 comparison. The standalone interface used five-fold evaluation with five training examples in the prompt and produced 99 polymer and 160 stability predictions.

On the original response scale, the standalone predictions gave $R^2 = -0.205$ for polymer HER and −2.701 for OER stability. Both values are below zero, that is, worse than predicting the mean response.

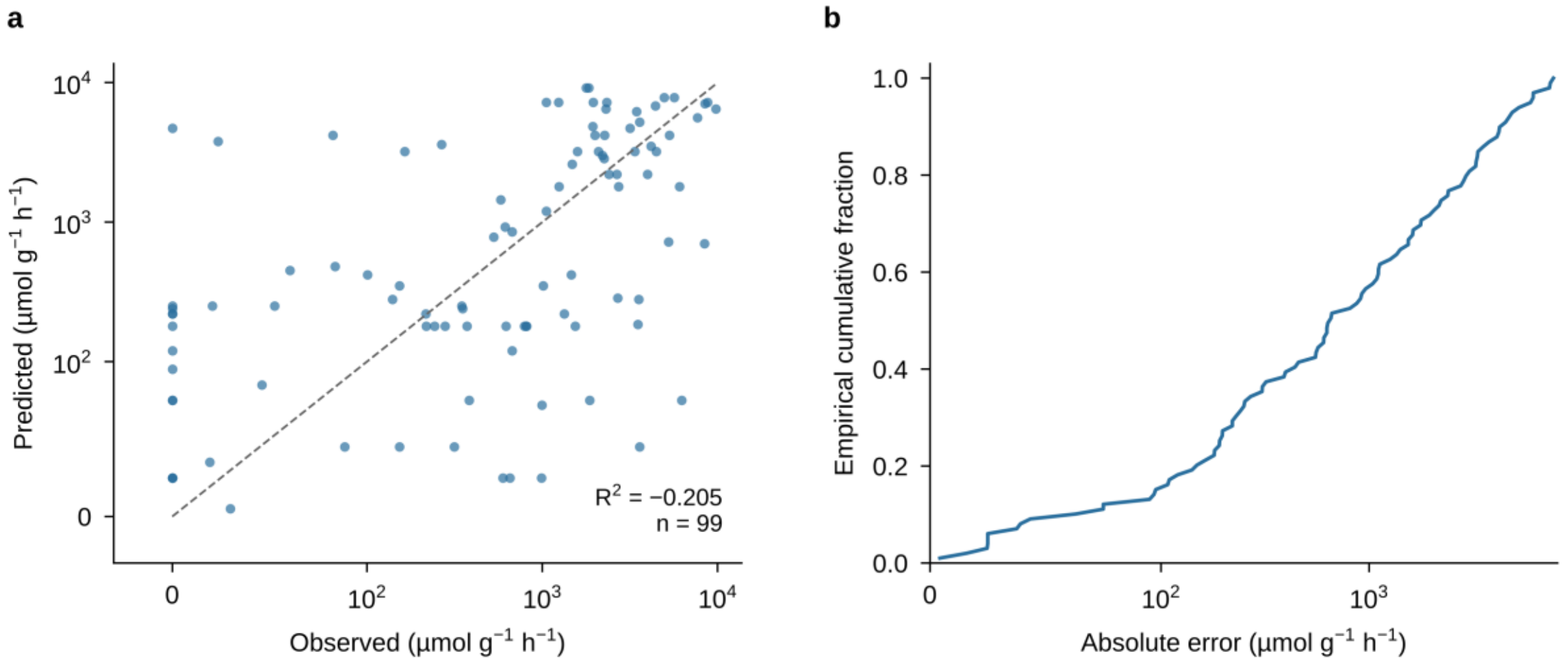

**Supplementary Fig. 2 | Logged GPT-5.5 standalone predictions for polymer HER. a**, Observed versus predicted responses for the standalone five-shot interface (99 predictions, five held-out folds); axes use a symmetric-log transformation (linear within ±100) to retain zero values. **b**, Empirical cumulative distribution of absolute errors on the original response scale, on a symmetric-log x axis. Supplementary Data 1 contains the logged predictions.

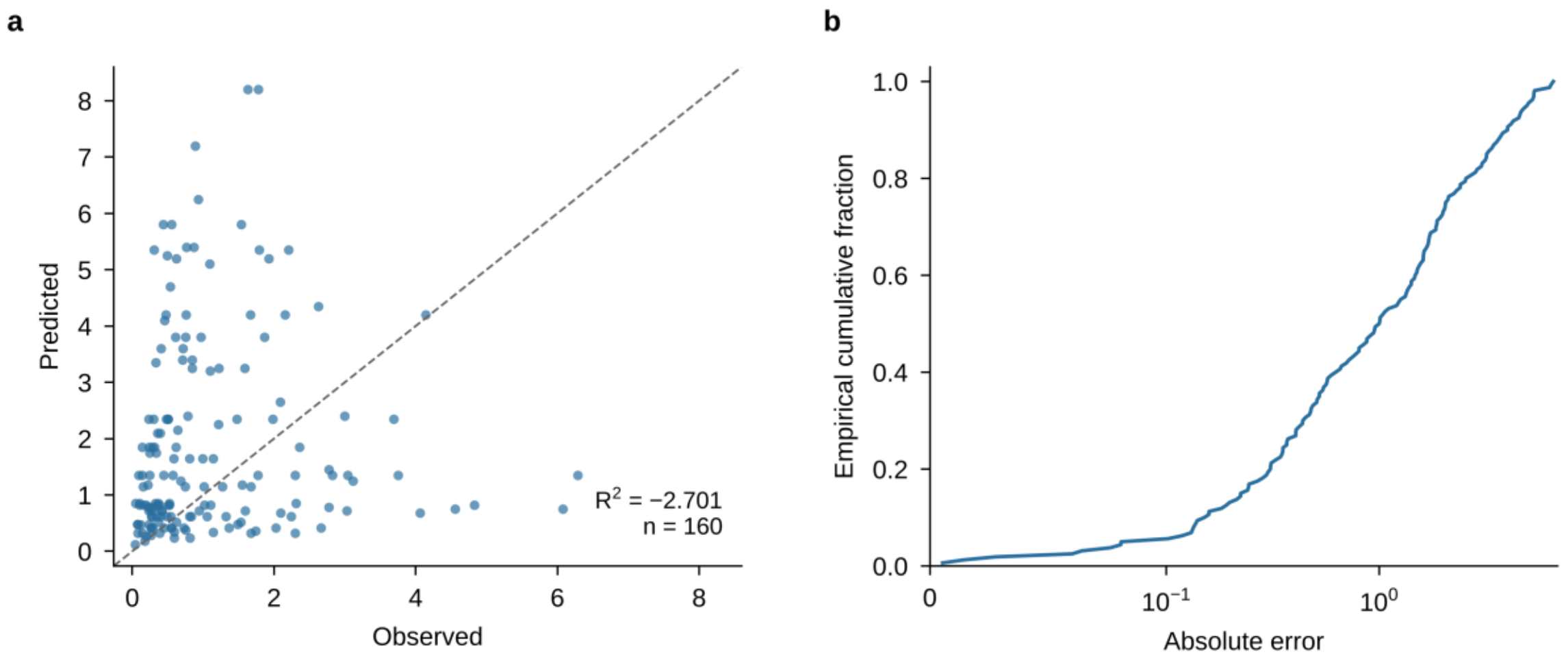

**Supplementary Fig. 3 | Logged GPT-5.5 standalone predictions for OER stability. a**, Observed versus predicted responses for the standalone interface (160 predictions, five held-out folds), in % Ru dissolution. **b**, Empirical cumulative distribution of absolute errors, on a symmetric-log x axis. Supplementary Data 1 contains the logged predictions.

**Supplementary Table 6 | Original-scale metrics of the logged GPT-5.5 standalone runs.** All metrics are recomputed from individual predictions. Polymer errors are in μmol $g^{-1}$ $h^{-1}$; stability errors are in % Ru dissolution.

| Task | Interface | n | $R^2$ | MAE | RMSE |
|---|---|---|---|---|---|
| Polymer HER | Standalone | 99 | −0.205 | 1,563.6 | 2,454.3 |
| OER stability | Standalone | 160 | −2.701 | 1.551 | 2.186 |

**Supplementary Note 8. Data versus representation: fixed-pipeline learning curves**

A model that performs poorly may need more data, or it may need a better representation. The learning-curve analysis asks whether further observations of the same kind would continue to improve a fixed pipeline. It uses an equally weighted ensemble of Extra Trees, random forest and gradient boosting rather than the Bayesian composition, so that changes in score reflect the data and not changes in model selection.

**Data and splits.** Polymer HER uses 99 records with 30 RDKit monomer and pair descriptors and five experimental descriptors (Pd content, measured optical gap, photoluminescence lifetime and transmittance in two media). OER stability uses 160 records with Ru dissolution (%) after 2 h, measured by inductively coupled plasma (ICP) analysis, as the target, 32 formula-fraction descriptors and five precursor descriptors; activity and post-test voltage are excluded from its inputs. Twenty random 80:20 splits were drawn for each task, giving training pools of 79 and 128 records and test sets of 20 and 32 records. Training subsets were nested within each split and evaluated on the same test records, at sizes 15–75 (polymer, steps of 10) and 20–125 (stability, steps of 15). Median imputation was refitted for every subset. Each estimator used 60 trees or boosting stages; the forests used a minimum leaf size of two and a feature fraction of 0.75, and gradient boosting used depth two with Huber loss. Repeat r used seed 2026 + r.

**Saturation fit.** Learning curves are commonly summarized by a parametric fit.[16] The fit uses the monotonic envelope of the mean held-out $R^2$, $\max_{k\leq j}\overline{R^2}(N_k)$, and the saturating form

$$g(N) = C - A\exp\left[-\frac{N - N_{\min}}{\tau}\right], \qquad \text{(S12)}$$

where C is the fitted ceiling, A the increase from the smallest training size and τ the saturation scale, with bounds $-1 \leq C \leq 0.99$, $10^{-5} \leq A \leq 5$ and $3 \leq \tau \leq 1{,}000$. For $A > 0$, the relative bottleneck index and half-improvement location are

$$B(N) = 1 - \exp\left[-\frac{N - N_{\min}}{\tau}\right], \qquad N^{*} = N_{\min} + \tau\ln 2. \qquad \text{(S13)}$$

The 90% intervals are the 5th and 95th percentiles over 2,000 bootstrap resamples of complete repeated curves (seeds 2026 and 12026), which preserves dependence between training sizes within a split. The splits reuse observations, so the intervals describe variation under the repeated-split construction rather than across independent datasets.

**Results.** The two tasks behave differently (Supplementary Fig. 4 and Supplementary Table 7). Polymer HER rises steadily to a fitted ceiling of 0.220 (90% interval 0.149–0.560), with half of the fitted improvement reached at $N^* \approx 32$ records. OER stability saturates at a ceiling of −0.031 (−0.089 to 0.077), an interval that includes zero. At the largest evaluated size, the fitted gain from ten more records is 0.0079 $R^2$ for polymer HER and 0.00007 for OER stability. These are extrapolations of the fitted curve, not measurements on new records. For this fixed tree ensemble, OER stability is limited by the representation and model family rather than by the number of observations, whereas polymer HER would still gain from more data of the same type. The scores are lower than the Bayesian-composition benchmark scores in main-text Fig. 3a,b because the pipeline, the descriptors and the evaluation protocol differ; the index describes saturation within this pipeline and is not a bound on SPARC performance.

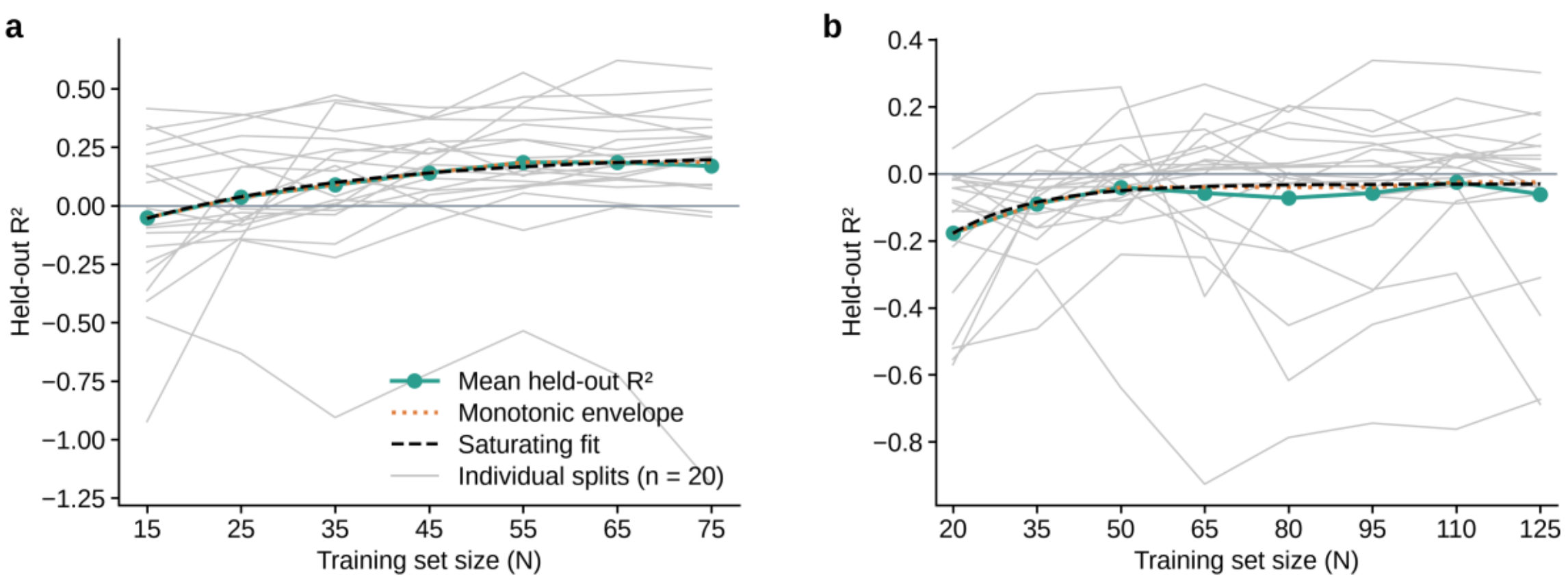

**Supplementary Fig. 4 | Split-level learning curves underlying the saturation diagnostic. a**, Polymer HER (99 records, seven training sizes). **b**, OER stability (160 records, eight training sizes). Thin grey lines show all 20 repeated 80:20 splits, each evaluating nested training subsets on a fixed test set. Teal points and lines show the mean held-out $R^2$ at each size, orange dotted lines its monotonic envelope and black dashed lines the fitted saturating response. Grey horizontal lines mark $R^2 = 0$. Every split-level value is shown, including negative values. Supplementary Table 7 gives the fitted parameters with 90% bootstrap intervals; Supplementary Data 2 gives all 300 split-level values.

**Supplementary Table 7 | Fixed-pipeline learning-curve parameters.** Brackets give 90% percentile bootstrap intervals from 2,000 resamples of the 20 complete curves. The gain in the last row is $g(N_{max} + 10) - g(N_{max})$ on the $R^2$ scale. The fitted amplitudes A are 0.274 (polymer HER) and 0.147 (OER stability).

| Quantity | Polymer HER | OER stability |
|---|---|---|
| Total records | 99 | 160 |
| Training pool / test records per split | 79 / 20 | 128 / 32 |
| Repeated splits | 20 | 20 |
| Evaluated training sizes | 15–75 | 20–125 |
| Fitted ceiling C | 0.220 [0.149, 0.560] | −0.031 [−0.089, 0.077] |
| Saturation scale τ (records) | 24.4 [14.2, 128.2] | 15.0 [5.8, 57.1] |
| Half-improvement location N* | 31.9 [24.8, 103.9] | 30.4 [24.0, 59.6] |
| Fitted gain for ten more records | 0.0079 | 0.00007 |

# Part III. From prediction to decisions

## Supplementary Note 9. Retrospective finite-pool acquisition

**Protocol.** In a retrospective finite-pool study, every candidate has a recorded outcome, but outcomes are hidden until the candidate is selected. Each search started from a fixed set of labelled records, identical for every strategy compared on a task. At each step the predictor was refitted on the labelled records, every unlabelled candidate was scored, and the highest-scoring candidate was revealed and added to the labelled set. One candidate was acquired per step, so each trajectory is fully sequential. SPARC used the heterogeneous bootstrap ensemble described in the main-text Methods with the trust-modulated confidence bound (equation S9, $\beta = 1.96$). The Gaussian-process comparator used a constant × Matérn-5/2 kernel with a white-noise term, response normalization and two optimizer restarts; the random-forest comparator used 300 trees, with the standard deviation across trees as its uncertainty. Both comparators ranked candidates with the same confidence bound and β, without the trust term ($T = 1$).

**Endpoint.** Discovery is the number of evaluations, including the initial ones, after which the best observed value first lies within 0.1% of the pool optimum. A single trajectory from one initialization is a case study; comparing strategies requires repeating the search over several initial sets and reporting the distribution of discovery counts.

## Supplementary Note 10. Controlled extrapolation: split definitions and uncertainty response

Main-text Fig. 4 asks whether the uncertainty of the predictor grows when it is asked about conditions outside the measured range, and whether batch selection under that uncertainty still finds good candidates.

**Split definitions.** The splits were defined from the inputs. For OER activity (462 records), cobalt levels 1–3 form the initial region, referred to here as the in-distribution (ID) set (406 records), and levels 4–7 form the extrapolation pool, or out-of-distribution (OOD) set (56 records). For formate oxidation (306 records), the 70th percentile of the first principal component (PC1) of the inputs separates 214 ID from 92 OOD records; an 80th-percentile sensitivity split gives 245 and 61. Both definitions use inputs only. The principal components are computed from all input rows without responses, so the formate-oxidation split is transductive in the inputs but not in the labels. All model fitting and uncertainty calibration use ID responses only.

**Predictor.** The predictor mixes three Gaussian-process experts: an isotropic Matérn-5/2 kernel, an automatic relevance determination (ARD) radial-basis-function kernel, and a sparse-prior GP (Supplementary Table 3). Each expert learns its observation noise through a white-noise kernel, with a numerical jitter of $10^{-6}$. Expert weights are the softmax of the mean pointwise three-fold cross-validated Gaussian log likelihood. For weights $\pi_m$, expert means $\mu_m(x)$ and predictive variances $v_m(x)$, the mixture moments are

$$\mu(x) = \sum_m \pi_m \, \mu_m(x), \qquad u_{\text{mix}}^2(x) = \sum_m \pi_m \, \{v_m(x) + [\mu_m(x) - \mu(x)]^2\}, \qquad \text{(S14)}$$

so the uncertainty contains both the variance within experts and their disagreement. Held-out mixture predictions for the training records, obtained by cross-fitting, give the residual scale $u_{res}$ and the held-out ID uncertainty used below.

**Uncertainty ratio.** The ratio of mean OOD to mean ID predictive standard deviation summarizes how much uncertainty grows outside the training support. Its value depends on the denominator. In-sample standard deviations are evaluated at the training records; held-out standard deviations are evaluated out of fold. Supplementary Table 8 therefore reports the ratio with three denominators. Replacing the in-sample denominator with the held-out ID standard deviation changed the OER-activity ratio only slightly (1.754 to 1.731) but reduced the formate-oxidation ratio more (1.266 to 1.129 at 70%; 1.288 to 1.138 at 80%). The main text reports the in-sample version.

**Calibrated variant.** A calibrated variant of the mixture was also evaluated. It multiplies the predictive standard deviation by a smooth support factor between one and four, built from three distances referenced to ID held-out 95th percentiles (standardized mean nearest-neighbour distance, Ledoit–Wolf shrinkage Mahalanobis distance[17] and distance to the convex hull in principal-component space, with the hull term capped), and by a single scale c learned from ID held-out residuals:

$$c_{\text{raw}}^2 = \frac{1}{n_{\text{ID}}} \sum_{i \in \text{ID}} \frac{(y_i - \mu_i)^2}{s_i^2}, \qquad c = \text{clip}(c_{\text{raw}}, 0.25, 4). \qquad \text{(S15)}$$

With calibration, the OER-activity ratio rose to 2.904 and nominal 90% intervals covered all 56 OOD observations, while $R^2$ was −7.743: the uncertainty was conservative, but the mean extrapolated poorly. For formate oxidation at the 70th percentile, the ratio was 1.165, coverage 0.978 and $R^2$ 0.520 (Supplementary Fig. 5). A large ratio is therefore not, by itself, evidence of good calibration; it must be read together with error and coverage.

**Group hold-out.** Leave-one-cobalt-level-out (seven OER-activity groups) and leave-one-PC1-quintile-out (five formate-oxidation groups) evaluations give 12 further checks (Supplementary Fig. 6). Their mean ratios were 1.251 (OER activity) and 1.446 (formate oxidation), with mean 90% coverage of 0.920 and 0.921. The calibrated ratio is U-shaped across groups: it is largest at both ends of the range (2.19 and 1.41 for cobalt levels 1 and 7; 3.23 and 1.26 for PC1 quintiles 1 and 5) and close to or below one for interior groups (0.98 for cobalt level 2; 0.78 for PC1 quintile 4), as expected if uncertainty responds to distance from the training support in either direction. One OER-activity cobalt level contains a single record and gives an especially imprecise coverage estimate.

**Batch selection.** Four rounds selected complete batches of 23 (formate oxidation) or 14 (OER activity) candidates by trust-modulated Thompson sampling (equation S10), with an independent standard-normal perturbation for each candidate. Each batch was fixed before its outcomes were revealed, and all experts and weights were refitted after each batch. The batch hit rate is the fraction of each batch in the top quartile of the pool, reported with a Wald interval and the 25% random expectation, and precision@K is the cumulative precision against the true top set of the same size.

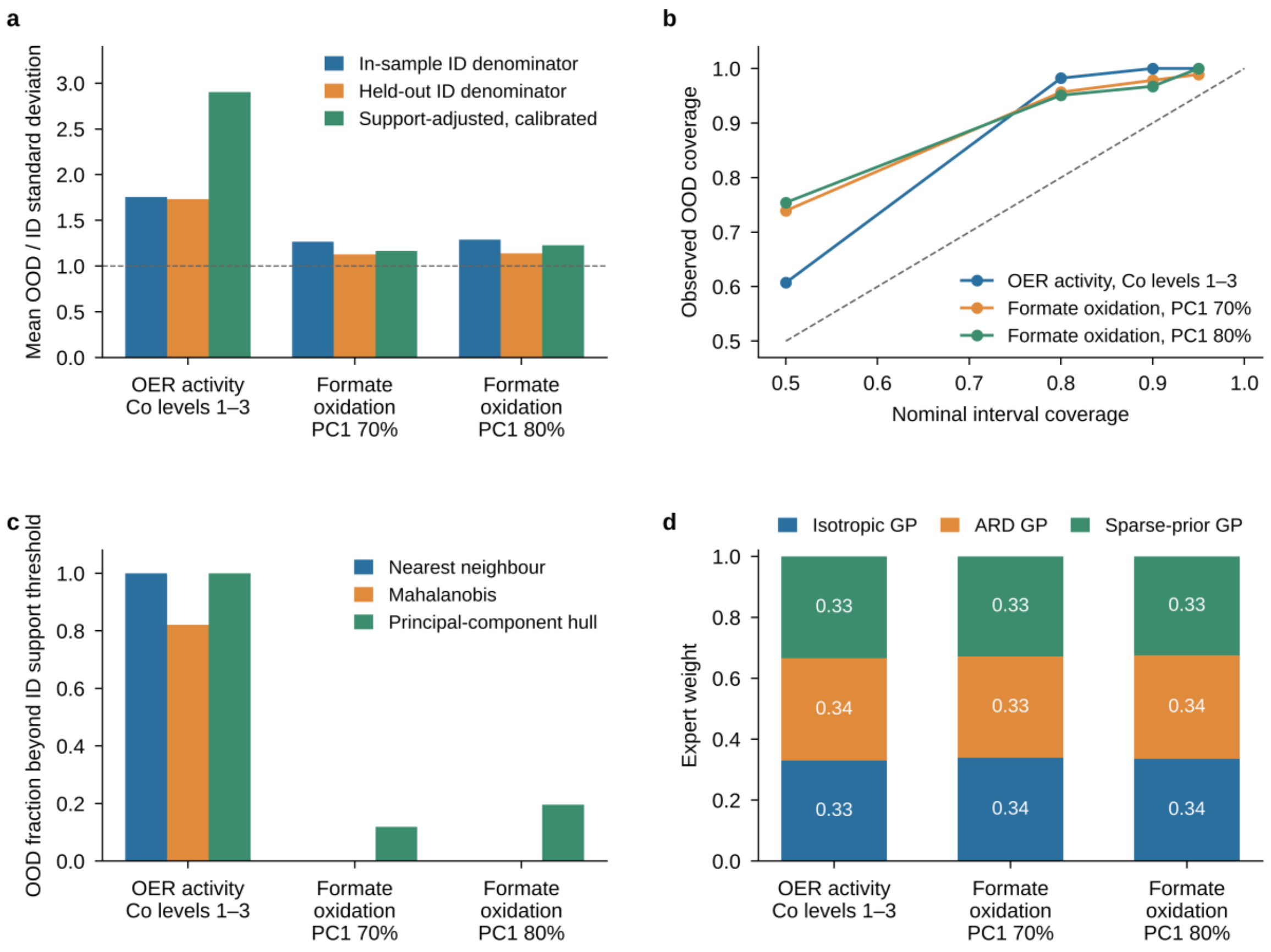

**Supplementary Fig. 5 | Calibration and feature support under defined distribution shifts. a**, OOD/ID predictive-standard-deviation ratios with an in-sample ID denominator, a held-out (out-of-fold) denominator and the support-adjusted calibrated model. **b**, Empirical OOD interval coverage at four nominal levels. **c**, Fractions of OOD records beyond ID support thresholds for nearest-neighbour, shrinkage Mahalanobis and principal-component hull distances. **d**, Expert weights learned from ID cross-fitted likelihoods; the third expert was a sparse-prior GP. The OER-activity split has 406 ID and 56 OOD records; the formate-oxidation splits have 214 and 92 (PC1 70th percentile) and 245 and 61 (80th percentile). OOD labels are not used in fitting or calibration. Supplementary Data 3 gives all metrics.

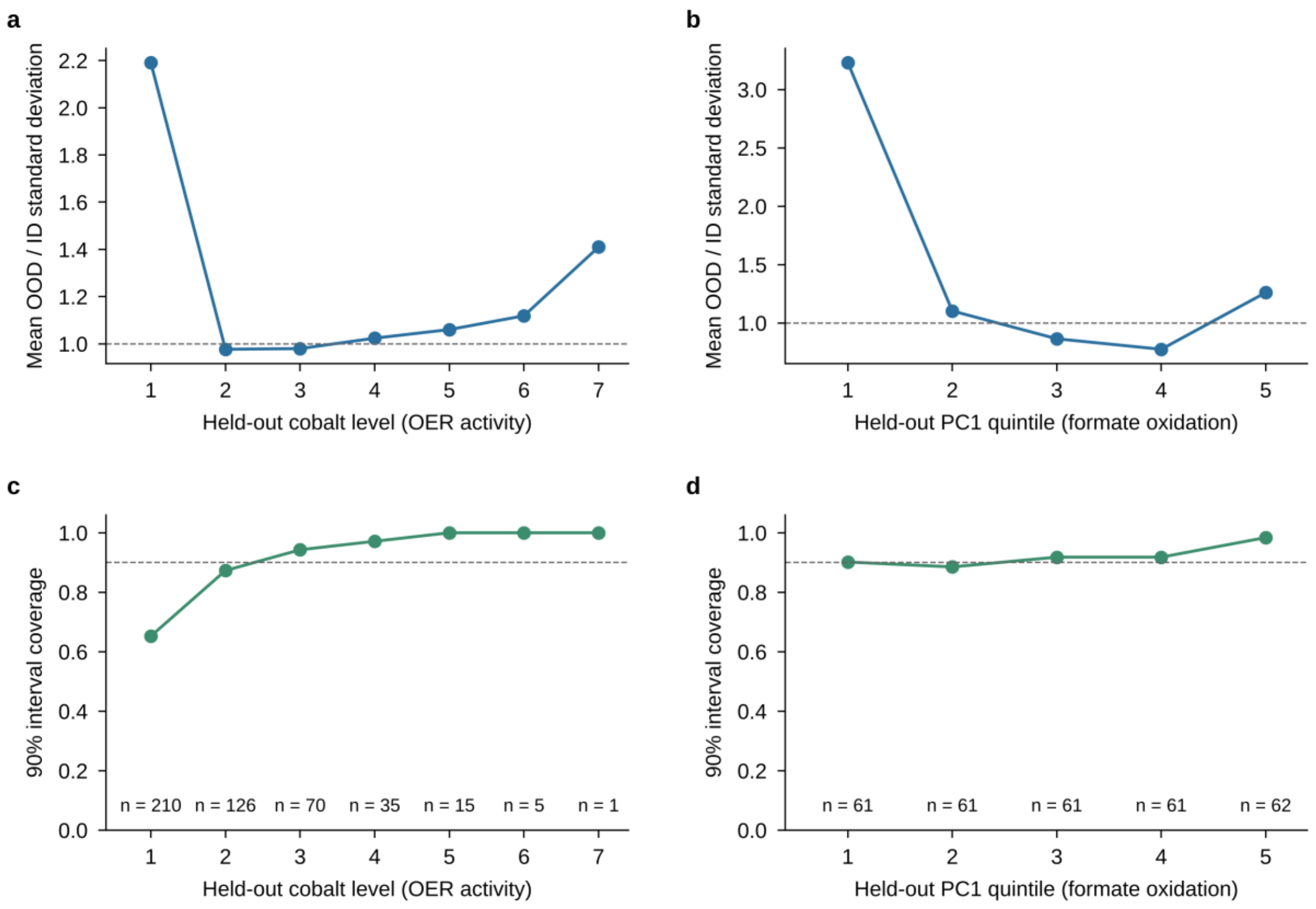

**Supplementary Fig. 6 | Leave-one-group-out sensitivity of OOD uncertainty. a,b**, Uncertainty ratios for each held-out cobalt level (OER activity) or PC1 quintile (formate oxidation). **c,d**, Coverage of nominal 90% intervals, with group sizes annotated. Each group is held out once and the remaining groups are used for training. The one-record cobalt level gives an imprecise coverage estimate. Lines connect group summaries and do not imply a continuous intervention.

**Supplementary Table 8 | Dependence of the OOD/ID uncertainty ratio on its denominator.** In-sample: mean ID training-record standard deviation (reported in the main text). Held-out: mean ID out-of-fold standard deviation. Calibrated: support-adjusted model with the held-out denominator (equation S15). Coverage and $R^2$ refer to the calibrated model on the OOD records.

| Split | ID / OOD records | In-sample | Held-out | Calibrated | 90% coverage | $R^2$ |
|---|---|---|---|---|---|---|
| OER activity, cobalt levels 4–7 | 406 / 56 | 1.754 | 1.731 | 2.904 | 1.000 | −7.743 |
| Formate oxidation, PC1 70th percentile | 214 / 92 | 1.266 | 1.129 | 1.165 | 0.978 | 0.520 |
| Formate oxidation, PC1 80th percentile | 245 / 61 | 1.288 | 1.138 | 1.227 | 0.967 | 0.536 |

## Part IV. Laboratory validation

### Supplementary Note 11. Automated OER campaign and its prospective prediction record

**Platform and design space.** The alkaline OER campaign used an automated electrodeposition and electrochemical characterization workflow built on the materials-acceleration and AMPERE-2 platforms.[3,18] The initial record contains 125 experiments spanning seven transition-metal cations with chloride and sulfate counter-anions and five synthesis variables: deposition temperature, deposition current (mA), deposition time, complexing-agent volume and precursor pH. The response is the overpotential required to reach 10 mA $cm^{-2}$ in linear-sweep voltammetry, in mV; lower is better.

**Synthesis and electrochemical measurement.** Metal chloride and sulfate precursors were dissolved in deionized water to give 1.2 M stock solutions, which were prepared fresh each day. A liquid-handling robot (Opentrons OT-2) dispensed the stocks and citric acid, the only complexing agent used, into the reactor wells according to each recipe and homogenized the bath by repeated aspiration and dispensing; the bath pH and temperature were recorded. The deposition temperature was set with the temperature control of the platform.[18] Films were deposited galvanostatically onto Ni foil (0.05 mm thick; cleaned by sonication in deionized water and in isopropanol for 20 min each) in a three-electrode configuration with an Ag/AgCl (1 M KCl) reference electrode and a Pt-wire counter electrode, using a potentiostat (BioLogic VSP-3e) controlled through a Python interface. The Ni foil was sealed against the bottom of each well by an O-ring, which defined a geometric electrode area of 0.283 $cm^2$. The reference and counter electrodes were cleaned in 1 M HCl and deionized water before each deposition. OER activity was measured in 1 M KOH, used as received without Fe purification, by linear-sweep voltammetry from 0 to 1.0 V versus the reference at 5 mV $s^{-1}$. Potentials were converted to the reversible hydrogen electrode scale ($E_{RHE} = E_{Ag/AgCl} + 0.235$ V $+ 0.059$ pH) and corrected for the series resistance obtained from impedance spectroscopy. The overpotential was read from the corrected voltammogram at 10 mA $cm^{-2}$ of geometric area.

**Recommendation endpoint.** Eleven conditions were recommended with the bootstrap ensemble (Supplementary Note 4) and the trust-modulated lower confidence bound (minimization form of equation S9, $\beta = 1.96$). Greedy batch selection skipped any candidate whose Euclidean distance to an already selected condition was below $0.15\sqrt{d}$ after standardizing the d inputs. The recommendations were fixed before measurement. Eight of the eleven reached overpotentials below 300 mV, with a minimum of 242 mV, compared with 18 of the 125 initial experiments. This compares the yields of two sets and is not a randomized estimate of an acquisition effect, because the recommended set was chosen to be promising.

**Prospective prediction record.** Because the recommendations were fixed before measurement, their predictions form a prospective test of the model (Supplementary Fig. 7). Over the 11 recommended conditions, the MAE between predicted and measured overpotential was 26.7 mV.

The errors were not symmetric. The four recommendations predicted below 300 mV (249–297 mV) were measured at 242–295 mV, with absolute errors of 7–37 mV; they included the best result. The other seven were predicted at 309–324 mV and measured at 254–308 mV. For four of them the model was pessimistic by 35–55 mV, and all seven were measured at or below their predictions. The model therefore ordered the two groups correctly but compressed the predicted range of the second group, so that several conditions it considered average were among the better ones measured. A pessimistic bias in the best part of the space is the expected failure of a model trained on a record in which only 18 of 125 experiments lie below 300 mV, and it is the bias that new measurements in that region correct.

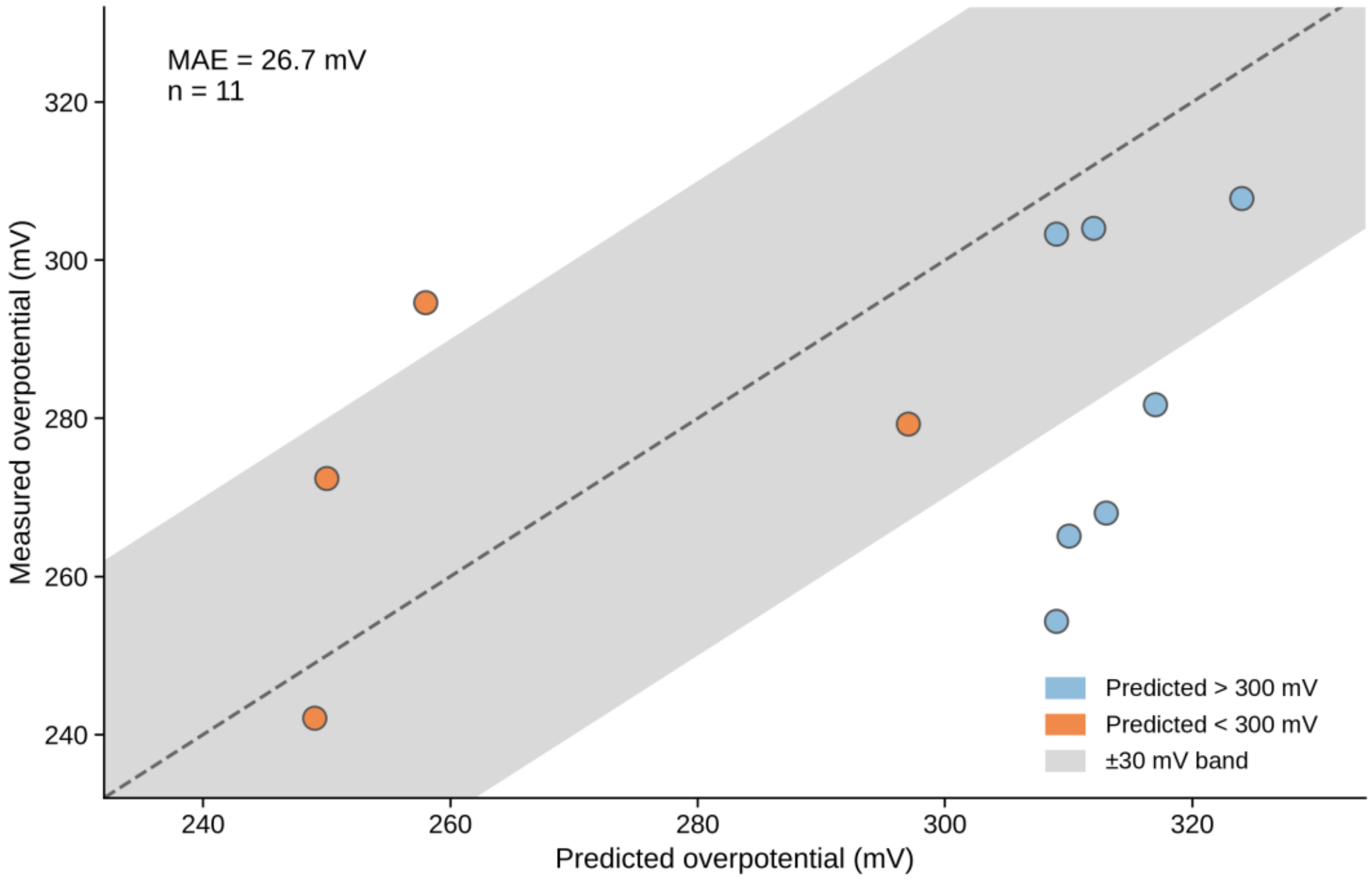

**Supplementary Fig. 7 | Prospective prediction record of the OER campaign.** Measured versus predicted overpotential for the 11 recommendations (MAE 26.7 mV); orange and blue, recommendations predicted below and above 300 mV; grey band, ±30 mV. Overpotentials are at 10 mA $cm^{-2}$.

## Supplementary Note 12. Measurement verification

**Diagnostics.** Automated experiments occasionally produce faulty measurements, and a model trained on them can be misled. SPARC therefore screens the training record with four diagnostics:

1. *Nonconformity.* The uncertainty-scaled absolute held-out residual is converted into a jackknife conformal $P$ value, the fraction of the other records with an equal or larger score (plus one, divided by n);[19] records with $P < 0.10$ are flagged.

2. *Influence.* The absolute difference between the prediction of a Ridge model ($\alpha = 1$) fitted to all records and the leave-one-out prediction of the model ensemble for the same record; the upper 10% are flagged.

3. *Isolation.* An Isolation Forest[20] with 200 trees and contamination 0.10.

4. *Mahalanobis distance.* Squared distance under a Ledoit–Wolf shrinkage covariance,[17] flagged when its nominal $\chi^2$ tail probability is below 0.05.

Trust (Supplementary Note 4) is reported alongside but is not counted as a fifth flag. Records with at least two flags are eligible for priority repetition, ordered by flag count, conformal $P$ value and influence. Controls are unflagged records matched to the priority records in standardized feature space. The thresholds are screening rules: a nominal tail probability does not establish calibration for dependent fitted residuals or non-Gaussian descriptors, and a flag does not prove that a measurement is wrong.

**Prospective repeats.** After the group assignments were fixed, seven priority measurements and eight controls were repeated (main-text Fig. 5f). The mean absolute original–repeat difference was 76 mV for priority records and 27 mV for controls, but the medians were similar (36.3 and 32.5 mV). Three priority differences exceeded 130 mV, whereas every control difference was below 42 mV. The separation lies in the tail of large deviations rather than in a shift of the typical deviation, which is the behaviour expected of a screen for occasional gross errors. These are descriptive summaries; with seven and eight records, no significance test is implied.

**Structure of the diagnostic record.** Among the 125 campaign records, 82 had no flag, 35 one flag and eight at least two (Supplementary Fig. 8). Nonconformity and influence provide different information: several records with large residuals had small influence, and the converse.

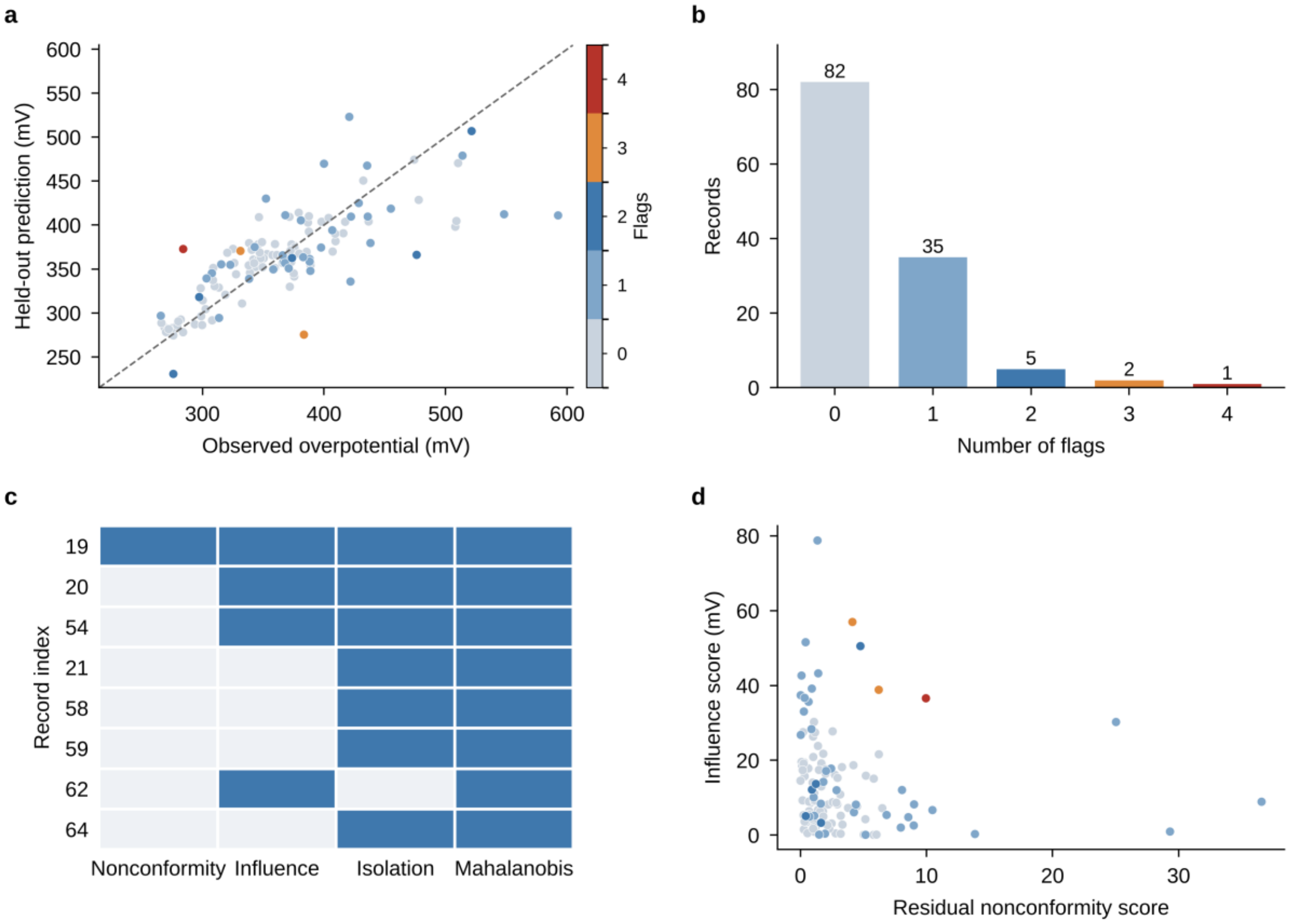

**Supplementary Fig. 8 | Diagnostic structure of the OER campaign record. a**, Observed versus held-out predicted overpotential for 125 records, coloured by total flags. **b**, Numbers of records with zero to four flags. **c**, Individual diagnostics for the eight records with at least two flags. **d**, Residual nonconformity versus the influence score (full-fit Ridge prediction minus the leave-one-out prediction of the model ensemble). Flags prioritize review and do not confirm erroneous measurements. Supplementary Data 4 contains the scores and flags.

# Part V. Reproducibility

## Supplementary Note 13. Software, settings and data

**Software.** SPARC is implemented in Python 3.12 with NumPy, SciPy, pandas, scikit-learn 1.6.1, NumPyro (JAX backend) with NUTS sampling[12] and SHAP; XGBoost is optional. The implementation is available at https://github.com/SissiFeng/SPARC. The package exposes one entry point per analysis in `sparc.figure_data`, each of which reads a JSON configuration (examples are supplied with the package) and writes the numerical source data of a figure. The workflows used in this Supplementary Information are: `ablation` (Supplementary Note 6), `finite_pool` (Supplementary Note 9), `extrapolation`, `pc_split` and `leave_group_out` (Supplementary Note 10) and `verification` (Supplementary Note 12). Every workflow records its configuration, package versions and random seeds with its output. Default settings are listed in Supplementary Tables 4 and 5.

**Reported results.** The plotting scripts read the stored numerical outputs and do not rerun model training or language-model inference. The workflows above rerun the corresponding analyses; Supplementary Table 3 lists the two settings in which the archived software differs from the reported analyses.

**Supplementary Data.** The following files accompany this Supplementary Information.

| File | Content | Used in (Supplementary Fig./Table) |
|---|---|---|
| Supplementary Data 1 | Logged GPT-5.5 standalone predictions: 259 records, two tasks | Figs 2, 3; Table 6 |
| Supplementary Data 2 | Repeated learning curves: 300 task–repeat–size rows | Fig. 4; Table 7 |
| Supplementary Data 3 | Calibrated OOD and group hold-out metrics: 15 split summaries | Figs 5, 6; Table 8 |
| Supplementary Data 4 | OER diagnostic scores and flags: 125 records | Fig. 8 |

## Supplementary References